%% file: main.tex
\documentclass[sigconf]{acmart}

\usepackage{makecell}
\usepackage{booktabs}
\usepackage{multirow}
\usepackage{tikz}
\usetikzlibrary{arrows.meta,calc,positioning}

\definecolor{stroke1}{RGB}{37,116,169}

\AtBeginDocument{%
  }

\copyrightyear{2026}
\acmYear{2026}
\setcopyright{cc}
\setcctype{by}
\acmDOI{10.1145/3816483.3816622}
\acmISBN{979-8-4007-2348-3/2026/06}

\acmConference[EASE '26]{International Conference on Evaluation and Assessment in Software Engineering}{June 09--12, 2026}{Glasgow, United Kingdom}
\acmBooktitle{International Conference on Evaluation and Assessment in Software Engineering (EASE '26), June 09--12, 2026, Glasgow, United Kingdom}

\begin{document}

\title[Evaluating Faithful LLM-Generated Failure Explanations]{From Program Slices to Causal Clarity: Evaluating Faithful, Actionable LLM-Generated Failure Explanations via Context Partitioning and LLM-as-a-Judge}

\author{Julius Porbeck}
\orcid{0009-0007-7998-0951}
\email{julius.porbeck@student.hpi.de}
\affiliation{%
  \institution{Hasso Plattner Institute, University of Potsdam}
  \city{Potsdam}
  \country{Germany}}

\author{Christian Medeiros Adriano}
\orcid{0000-0003-2588-9937}
\email{christian.adriano@hpi.de}
\affiliation{%
  \institution{Hasso Plattner Institute, University of Potsdam}
  \city{Potsdam}
  \country{Germany}}

\author{Holger Giese}
\orcid{0000-0002-4723-730X}
\email{holger.giese@hpi.de}
\affiliation{%
  \institution{Hasso Plattner Institute, University of Potsdam}
  \city{Potsdam}
  \country{Germany}}

\renewcommand{\shortauthors}{Porbeck et al.}

\begin{abstract}
Large language model (LLM)-based debugging systems can generate failure explanations, but these explanations may be incomplete or incorrect. Misleading explanations are harmful for downstream tasks (e.g., bug triage, bug fixing). We investigate how explanation quality is affected by various LLM context configurations. Existing work predominantly treats LLM-generated failure explanations as an ad hoc by-product of debugging or repair workflows, using generic prompting over undifferentiated artifacts such as code, tests, and error messages rather than targeting explanations as a first-class output with dedicated quality assessment. Consequently, existing approaches provide limited support for assessing whether these explanations capture the underlying fault–error–failure mechanism and for actionable next steps, and most techniques instead prioritize task success (e.g., patch correctness or review quality) over the explicit causal explanation quality. We systematically vary the debugging information to study how distinct context compositions affect the quality of LLM-generated failure explanations. Across 93 context configurations on real bugs and three economically viable models (\texttt{gpt\allowbreak-5-mini}, \texttt{DeepSeek\allowbreak-V3.2}, and \texttt{Grok\allowbreak-4.1-fast}), we evaluate explanations with six criteria and validate the LLM-as-a-judge scores against human ratings in a user study.
Our results indicate that explanation quality is causally affected by context composition. Evidence-rich, failure-specific artifacts improve causal and action-oriented quality, whereas overly large contexts tend to yield vague explanations. Higher explanation-score quartiles are associated with higher downstream repair pass rates and, for some models, with fixes that are closer to the reference minimal fixes. In contrast, low-score quartiles can even underperform the no-explanation baseline. Reproduction package is publicly available.\footnote{\href{https://doi.org/10.5281/zenodo.19556994}{\nolinkurl{https://doi.org/10.5281/zenodo.19556994}}}


\end{abstract}



\maketitle


\input{1_introduction/introduction}
\input{2_foundations_related_work/foundations_related_work}

\input{3_problem_formulation_and_rqs/problem_formulation_and_rqs}
\input{4_methodology_and_experimental_design/methodology_and_experimental_design}
\input{5_results/results}
\input{6_discussion_and_implications/discussion_and_implications}
\input{7_threats_to_validity/threats_to_validity}

\input{8_conclusion_and_future_work/conclusion_and_future_work}

\bibliographystyle{ACM-Reference-Format}
\bibliography{lit}

\end{document}

%% file: 1_introduction/introduction.tex
\section{Introduction}
\textbf{Context} -- Software failures demand a precise understanding. While large language models (LLMs) can expedite the explanation of various types of software failures, these can still be incomplete or vague. This can mislead downstream tasks such as bug triage and bug fixing, even when these tasks are automated. Hence, a lack of explanation quality more than an interpretability issue, it becomes a validation problem for the LLM-based bug-fixing pipeline.

\textbf{Current Limitations} -- Current work predominantly treats LLM-generated failure explanations as an ad hoc by-product of code review, debugging, or repair workflows. These workflows typically rely on generic prompting over entire undifferentiated artifacts such as code, tests, error messages, or pull-requests~\cite{Holloway2024ContextGranularityAPR,adhalsteinsson2025rethinking,mariotto2025fromassessment}. 
In practice, current LLM-based solutions have retained the automated program repair (APR)~\cite{Monperrus2014APRReview,Qi2015PlausiblePatches} approach by prioritizing task success (e.g., patch correctness or review quality) over the quality of the accompanying explanations.
Consequently, existing approaches provide limited support for assessing whether these explanations capture the underlying fault–error–failure mechanism and for actionable next steps.

\textbf{Insights} -- Explanation quality is partly subjective, yet it must capture causal knowledge about which program elements and conditions made the failure occur and what would need to change for it to disappear. We hypothesize that the reasons for incomplete or vague explanations arise from ad hoc context compositions that ignore trade-offs between criteria that can subjective (readability, conciseness), objective (faithfulness to artifacts), or actionable (conducive to valid fixes). In addition, conditioning LLMs on long token sequences -- such as undifferentiated free-text and large code dumps -- is known to dilute attention~\citep{an2025why,Liu2023LostInMiddle}.

\textbf{Approach} -- We systematically investigate how manipulating the context affects the \textit{quality}, \textit{faithfulness}, and \textit{actionability} of LLM-generated failure explanations~\cite{woodward1989causal}. We vary context composition using debugging artifacts across 93 configurations on real bugs from popular open-source projects and three economically viable models (\texttt{gpt\allowbreak-5-mini}, \texttt{DeepSeek\allowbreak-V3.2}, and \texttt{Grok\allowbreak-4.1-fast}), assess explanation quality along six criteria, and validate LLM-as-a-judge ratings against human ratings in a user study. This paper builds on and extends the first author's Master's thesis~\cite{porbeck2026program}.

\textbf{Results} -- The strongest gains come from executable, failure-specific artifacts such as \texttt{CODE} and \texttt{TEST}; in contrast, adding broad or prose-heavy context can make explanations vaguer. Higher explanation-score quartiles also show higher downstream repair pass rates and, for some models, fixes closer to the reference minimal fixes, while low explanation-score quartiles can underperform with respect to the no-explanation baseline.

\textbf{Contributions} are threefold:
\begingroup
\setlength{\topsep}{0pt}
\begin{enumerate}
    \item \textbf{A context-module catalog and context-partitioning pipeline.} We define context as composable modules and implement a pipeline that generates and evaluates explanations and downstream fixes under varied context compositions.
    \item \textbf{A failure-explanation-specific evaluation framework with six quality criteria.} We operationalize explanation quality as a multi-dimensional construct and validate judge-based criteria against human judgments in a user study.
    \item \textbf{An empirical study of context partitioning and downstream repair precision.} We quantify how context composition shapes explanation quality and relate explanation quality to downstream repair success and repair precision (similarity to a reference minimal fix).
\end{enumerate}
\endgroup



%% file: 2_foundations_related_work/foundations_related_work.tex
\section{Foundations and Related Work}

 

The automation of failure explanations overlaps with dependability theory, cognitive debugging, program analysis, and, more recently, LLM context management. We review how these areas define fault–error–failure chains, how developers actually search for causes, how program slicing and context partitioning can isolate failure-relevant evidence, and how existing evaluation work falls short of a failure-explanation–specific framework. Two trends stand out: (1) context work is shifting from “more is better” toward selective, multi-source workflows for LLM-based software engineering, and (2) evaluation work increasingly treats developer-facing text as a multi-dimensional object with distinct criteria for usefulness, grounding, and surface factors (readability and brevity). However, these trends are rarely unified for failure explanations, leaving a gap of a dedicated quality framework paired with a principled method for composing context.

\subsection{Explanation Failure Role in Debugging}

A failure explanation answers \textit{“Why did this failure happen?”} by linking an observed failure (e.g., a failing test) to the program elements and conditions that caused it, including relevant inputs and configurations ~\cite{adriano2022microtasking,widyasari2024demystifying}. It should go beyond symptoms by identifying what must change for the failure to disappear, recognizing that a fault becomes an observable failure only if it is executed, produces an incorrect state, and that state propagates to the output ~\cite{Avizienis2004Taxonomy,VoasMiller1995Testability}. Because developers rely on multiple sources—code, inputs, and environment configurations—explanations often arise from ad hoc mental models that can be technically correct yet unhelpful for downstream tasks~\cite{layman2013debugging,li2026grounded}.

Existing work has begun to add natural-language explanations to fault localization, but with different goals and granularity than ours. Yan et al. \cite{yan2024better} combines static analysis and LLMs to generate explainable, fault-localization reports that explain crashes and evaluates them primarily through user satisfaction, without a multi-dimensional explanation-quality framework or explicit fault–error–failure chains. Widyasari et al.~\cite{widyasari2024demystifying} and Salmon et al.~\cite{salmon2025debugging} study LLM-generated explanations using criteria such as usefulness, correctness, and verbosity, but focus on step-by-step reasoning or novice error feedback for single-artifact programs rather than system-level, multi-source failure explanations. Our work contributes to this literature by adding software failures with distributed evidence, judged against a set of idealized quality criteria and causal-effect on bug-fix success.

\subsection{Context Windows and Context Partitioning}
For an LLM, the \emph{context} is the sequence of tokens provided at inference time, and the \emph{context window} is the model-specific limit on the length of this sequence~\cite{wang2024beyond,an2025why}. Because the model conditions its output on this context, what we omit, include, and order has a causal-effect\footnote{Assuming that manipulating context corresponds to an intervention~\cite{woodward1989causal}.} on the resulting outcome~\cite{guan2025the}. In LLM-based software engineering~\cite{rando2025longcodebench}, it might be tempting to assume that more context always enables better reasoning and to paste entire files, long traces, and supplementary documentation into a single prompt. However, likely due in part to how attention and learned inductive biases scale with length, long inputs can remain difficult even when they fit within the nominal window~\cite{an2025why}. Empirically, these limitations give rise to a \textit{lost-in-the-middle} effect: information in the middle of long inputs is used less reliably than information near the beginning or end~\cite{Liu2023LostInMiddle}.


Context partitioning addresses this by treating the prompt as a composed object rather than a single text blob. We use it to split large information sources into parts and construct prompts that maximize semantic signal under a context budget by carefully selecting, structuring, and composing failure-relevant evidence. For failure explanations, quality depends on how the debugging evidence is selected and placed within the context window, and program slicing can act as a context-selection mechanism by extracting statements likely to influence a failure observation and augmenting them with minimal surrounding code~\cite{soremekun2021locating}. This reduces irrelevant tokens and makes relevant evidence easier for the model to attend to, without assuming it will reliably find relevance in a noisy prompt. Prior work on LLM-driven program repair reports a trade-off between broad and targeted contexts: broader contexts can improve aggregate success but risk semantic dilution, whereas more focused contexts can improve precision by concentrating attention on relevant snippets \cite{Holloway2024ContextGranularityAPR}. Similarly, workflow-level approaches to LLM-assisted code review that combine multiple artifacts in a retrieval-augmented setup motivate treating context as an object assembled from modules, which we adopt for the context-partitioning pipeline \cite{adhalsteinsson2025rethinking}.

\subsection{Evaluating LLMs Beyond Task Success}

Automated program repair (APR) systems propose patches, apply them, and accept patches that pass test suite~\cite{Monperrus2014APRReview}. This loop is operationally efficient, but passing tests does not guarantee semantic correctness, because APR systems can produce \emph{plausible} patches that overfit the test~\cite{Qi2015PlausiblePatches}. If we claim that one explanation is better than another, we therefore need a way to measure quality beyond task success. Reference-based similarity metrics are a common starting point, but they are weak proxies for failure explanations, since correct explanations can differ substantially in wording and surface overlap can still miss or distort the underlying causal link. This motivates evaluation methods that assess faithfulness and usefulness more directly, especially for explanations that are meant to support debugging decisions.

LLM-as-a-judge evaluation uses an LLM to score or rank candidate outputs according to an explicit rubric, offering judgments at scale ~\cite{Zheng2023MTBench,Liu2023GEval}. However, judge models can be biased—for example, by favoring verbosity or applying criteria inconsistently—so they must be treated as measurement instruments with threats to validity. Conceptual safeguards include precise criteria and separation of judgment dimensions, randomized presentation to reduce position effects, calibration on examples, and consistency checks. Various work have investigated multi-dimensional quality frameworks for developer-facing artifacts~\cite{he2026llm}, for instance, DeepCRCEval introduces nine criteria for evaluating code review comments \cite{lu2025deepcrceval}, and an explanation taxonomy for fault prediction distinguishes Descriptive, Contextual, and Actionable explanations \cite{adejumo2025explainingcoderiskoss}. However, failure explanations are still overlooked regarding judgement criteria that evaluates both causal clarity (why the failure happened) and actionability (what to inspect or change). This is a methodological gap we aim to bridge.


%% file: 3_problem_formulation_and_rqs/problem_formulation_and_rqs.tex
\section{Problem and Research Questions}

\subsection{Task Definition and Setting}
We consider a localized, reproducible software failure triggered by a unit test.
The system has access to artifacts that are typically available during debugging, including source code, the failing test, the resulting error message, documentation, and program-analysis-derived evidence.
The desired output is a \emph{causal} explanation: an account of what program elements and conditions made the failure occur, and what would need to change for the failure to disappear.

We treat available evidence as composable \emph{context modules} and study how context composition influences explanation quality.
Concretely, let \(M\) denote the fixed set of modules available for each defect. It comprises four direct artifacts (\texttt{CODE}, \texttt{ERROR}, \texttt{TEST}, and \texttt{DOCSTRING}, where \texttt{DOCSTRING} may be empty) and four slicing-derived modules (\texttt{SLICE\_\allowbreak{}BLOCK}, \texttt{SLICE\_\allowbreak{}BACKWARD}, \texttt{SLICE\_\allowbreak{}FORWARD}, \texttt{SLICE\_\allowbreak{}UNION}).
A context configuration is a subset \(c \subseteq M\) (possibly all modules); \texttt{BASELINE} corresponds to \(c=M\).
Given a defect instance \(d\) and configuration \(c\), an LLM \(f_{\theta}\) generates a failure explanation \(e=f_{\theta}(d,c)\).
To operationalize downstream \emph{utility}, we additionally compare repairs generated (i) without an explanation and (ii) with the generated explanation appended, validating fixes by re-running the failing test and, for passing fixes, comparing against a defect-specific reference minimal fix.

\subsection{Explanation Quality as a Multi-Dimensional Construct}
Before we can optimize context, we must define what it means for an explanation to be of \emph{high quality}.
This is difficult because explanations are multi-dimensional and partly subjective: two explanations can be phrased very differently and still be correct, while a fluent explanation can contain subtle mistakes.
Unlike code generation, their semantic correctness cannot be automatically validated.

We therefore treat explanation quality as a \emph{multi-dimensional construct} and operationalize it via six binary criteria (C1--C6).
At a high level, the criteria capture (i) causal understanding (C2: correct problem identification; C3: coherent causal chain), (ii) actionability (C4: concrete, developer-usable next steps), (iii) grounding in the provided artifacts (C5), and (iv) surface factors (C1: readability; C6: brevity).
Because large-scale human evaluation is expensive, we use an automated \emph{LLM-as-a-judge} protocol for the judgment-based criteria and complement it with deterministic checks for criteria that can be evaluated directly.
We treat judge-based criteria as a measurement instrument that must be validated against human judgment.

For an explanation \(e\), let \(\mathbf{C}(e)\in\{0,1\}^{6}\) denote the per-criterion outcomes and let
\[
q(e)=\sum_{i=1}^{6} C_{i}(e)
\]
denote the total explanation quality score.
We analyze per-criterion rates and \(q(e)\) (e.g., via stratification into score quartiles) to relate context composition and intermediate explanation quality to downstream repair outcomes.

\subsection{Research Questions}
We address four research questions:
\begin{enumerate}
    \item \textbf{RQ1:} Do automated LLM-as-a-judge and humans produce similar evaluations of failure explanations? 
    \item \textbf{RQ2:} Are certain evaluation criteria more discriminative than others?
    \item \textbf{RQ3:} Is the quality of failure explanations sensitive to context prompting?
    \item \textbf{RQ4:} Do higher explanation-quality quartiles show generated fixes that better match a reference minimal fix (i.e., minimal, non-spurious code changes)?
\end{enumerate}  

To operationalize RQ1 (human vs.\ LLM-as-a-judge), we decompose it into four complementary agreement analyses (AQ1--AQ4), each answered by one results table or figure in Section~\ref{sec:results}.

%% file: 4_methodology_and_experimental_design/methodology_and_experimental_design.tex
\section{Methodology and Experimental Design}

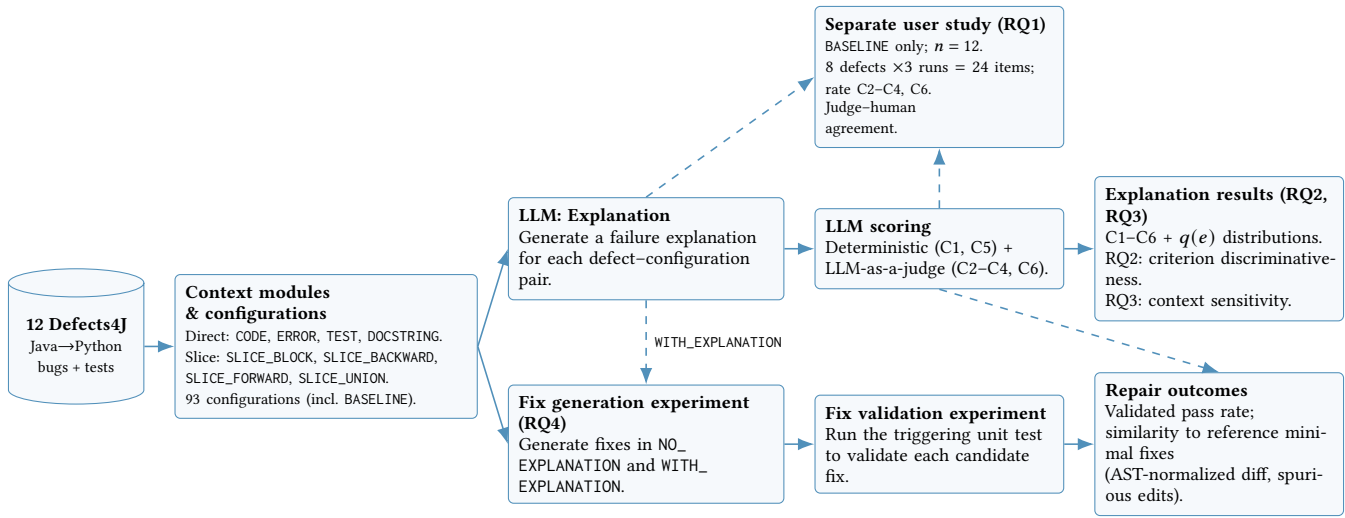
\begin{figure*}[t]
    \centering
    \usetikzlibrary{shapes.geometric}
    \begin{tikzpicture}[
        node distance=8mm and 4mm,
        overviewBox/.style={
            draw=stroke1!80,
            fill=stroke1!4,
            rounded corners=2pt,
            align=left,
            inner sep=4pt,
            font=\footnotesize,
            text width=0.19\linewidth
        },
        overviewBoxWide/.style={overviewBox, text width=0.21\linewidth},
        overviewBoxNarrow/.style={overviewBox, text width=0.17\linewidth},
        overviewDB/.style={
            draw=stroke1!80,
            fill=stroke1!4,
            cylinder,
            shape border rotate=90,
            aspect=0.25,
            minimum height=18mm,
            minimum width=18mm,
            align=center,
            inner sep=2pt,
            font=\footnotesize
        },
        overviewArrow/.style={-{Latex[length=2.2mm]}, line width=0.6pt, draw=stroke1!80},
        overviewDashed/.style={overviewArrow, dashed},
    ]
        \node[overviewDB] (dataset) {\textbf{12 Defects4J}\\{\scriptsize Java$\to$Python}\\{\scriptsize bugs + tests}};

        \node[overviewBoxWide, right=of dataset] (context) {%
            \textbf{Context modules\\ \& configurations}\\
            {\scriptsize Direct: \texttt{CODE}, \texttt{ERROR}, \texttt{TEST}, \texttt{DOCSTRING}.}\\
            {\scriptsize Slice: \texttt{SLICE\_\allowbreak{}BLOCK}, \texttt{SLICE\_\allowbreak{}BACKWARD}, \texttt{SLICE\_\allowbreak{}FORWARD}, \texttt{SLICE\_\allowbreak{}UNION}.}\\
            {\scriptsize 93 configurations (incl.\ \texttt{BASELINE}).}
        };

        \node[overviewBox, right=of context, yshift=13mm] (explain) {%
            \textbf{LLM: Explanation}\\
            Generate a failure explanation for each defect--configuration pair.
        };
        \node[overviewBoxNarrow, right=of explain] (rate) {%
            \textbf{LLM scoring}\\
            Deterministic (C1, C5) +\\
            LLM-as-a-judge (C2--C4, C6).
        };
        \node[overviewBoxNarrow, right=of rate] (quality) {%
            \textbf{Explanation results (RQ2, RQ3)}\\
            C1--C6 + \(q(e)\) distributions.\\
            RQ2: criterion discriminativeness.\\
            RQ3: context sensitivity.
        };

        \node[overviewBox, right=of context, yshift=-13mm] (fixgen) {%
            \textbf{Fix generation experiment (RQ4)}\\
            Generate fixes in \texttt{NO\_\allowbreak{}EXPLANATION} and \texttt{WITH\_\allowbreak{}EXPLANATION}.
        };
        \node[overviewBoxNarrow, right=of fixgen] (test) {%
            \textbf{Fix validation experiment}\\
            Run the triggering unit test to validate each candidate fix.
        };
        \node[overviewBoxNarrow, right=of test] (outcome) {%
            \textbf{Repair outcomes}\\
            Validated pass rate;\\
            similarity to reference minimal fixes\\
            (AST-normalized diff, spurious edits).
        };

        \node[overviewBoxNarrow, above=of rate] (human) {%
            \textbf{Separate user study (RQ1)}\\
            {\scriptsize \texttt{BASELINE} only; $n=12$.}\\
            {\scriptsize $8$ defects $\times 3$ runs $=24$ items; rate C2--C4, C6.}\\
            {\scriptsize Judge--human\\ agreement.}
        };

        \draw[overviewArrow] (dataset.east) -- (context.west);
        \draw[overviewArrow] (context.east) -- (explain.west);
        \draw[overviewArrow] (context.east) -- (fixgen.west);

        \draw[overviewArrow] (explain.east) -- (rate.west);
        \draw[overviewArrow] (rate.east) -- (quality.west);

        \draw[overviewDashed] (explain.north) -- (human.west);
        \draw[overviewDashed] (rate.north) -- (human.south);

        \draw[overviewDashed] (rate.south) -- (outcome.north);

        \draw[overviewArrow] (fixgen.east) -- (test.west);
        \draw[overviewArrow] (test.east) -- (outcome.west);

        \draw[overviewDashed] (explain.south) --
            node[midway, right, font=\scriptsize]{\texttt{WITH\_\allowbreak{}EXPLANATION}}
            (fixgen.north);
    \end{tikzpicture}
    \caption{Study overview: from 12 Defects4J defects translated from Java to Python (contamination control: Java \texttt{NO\_\allowbreak{}EXPLANATION} baseline; manual inspection), we vary context composition (8 modules; 93 configurations), generate failure explanations (3 models $\times$ 3 runs; temperature $=0$ where possible), score them via deterministic checks and LLM-as-a-judge, analyze criterion discriminativeness (RQ2) and context sensitivity (RQ3), validate judge ratings via a separate human user study (RQ1), and assess downstream utility via fix generation and fix validation with and without explanations (RQ4).}
    \Description{A two-lane overview diagram. A database-shaped node labeled ``12 Defects4J'' notes a Java-to-Python translation and includes bugs and tests. It feeds into a context-modules-and-configurations box listing eight context modules and noting 93 configurations. The top lane shows LLM explanation generation, automated scoring via deterministic checks and an LLM-as-a-judge rubric, and an explanation-results box labeled RQ2 and RQ3. A separate user-study box labeled RQ1 is connected via dashed arrows to indicate that humans rate \texttt{BASELINE} explanations and their ratings are compared against judge labels; another dashed arrow indicates that explanation scoring informs repair analyses. The bottom lane shows a fix-generation experiment (RQ4), a fix-validation step that reruns the triggering unit test, and repair outcomes including pass rate and similarity to reference minimal fixes. A dashed arrow from explanation generation to fix generation indicates the \texttt{WITH\_\allowbreak{}EXPLANATION} condition.}
    \label{fig:overview}
\end{figure*}

\subsection{Dataset and Defect Selection}
Figure~\ref{fig:overview} summarizes our study scenario: from a $12$-defect dataset, we validate LLM-as-a-judge ratings against human ratings via a separate user study (RQ1), score explanations via C1--C6 and analyze criterion discriminativeness (RQ2), vary context composition (RQ3), and assess downstream utility via fix generation and fix validation with and without explanations (RQ4).

We construct the dataset by selecting function-local Defects4J defects \citep{Just2014Defects4J} that are reproducible with a unit test in one to two steps; this filter reduces repair-scope confounds and keeps the 93-configuration, three-model experiment tractable.
To reduce contamination risk and direct memorization from pretraining, we manually ported each selected Java defect to Python and packaged it with its triggering unit test.
The porting preserved the localized failing behavior and adapted only language- and library-specific details needed to make the Python artifact executable.
We checked each translated defect by running the packaged triggering test and retained only cases where the expected failure reproduced in one to two steps.
The dataset contains $12$ translated Python defects.
As a contamination control, we additionally run a Java \texttt{NO\_\allowbreak{}EXPLANATION} baseline on the original Defects4J defects and assess Java fixes by manual inspection.
Across models, the Java \texttt{NO\_\allowbreak{}EXPLANATION} baseline achieves $72$--$78\%$ pass rate, whereas on the translated Python defects the same baseline achieves only $31$--$50\%$ validated pass rate, which is consistent with the translation process reducing contamination risk.

\subsection{Context Modules and Composition Strategy}
For each defect, we treat available evidence as composable \emph{context modules} and systematically vary which modules are provided to the model.
We use the fixed set \(M\) of eight modules introduced in Section~3: four direct artifacts (\texttt{CODE}, \texttt{ERROR}, \texttt{TEST}, \texttt{DOCSTRING}) and four slicing-derived modules (\texttt{SLICE\_BLOCK}, \texttt{SLICE\_BACKWARD}, \texttt{SLICE\_FORWARD}, \texttt{SLICE\_UNION}).
The derived \texttt{BASELINE} configuration contains all eight modules and serves as a reference point.

Our evaluation covers 93 unique context configurations, spanning isolated modules, pairwise and three-way combinations, and the \texttt{BASELINE} configuration.
This includes $8$ isolated configurations, $\binom{8}{2}=28$ pairwise configurations, $\binom{8}{3}=56$ three-way configurations, and \texttt{BASELINE}.
In the execution protocol, runs are organized into three batches (isolated, two-way, three-way), each including \texttt{BASELINE} as a reference point; thus each defect/run evaluates $(8+1) + (28+1) + (56+1) = 95$ configuration instances, with \texttt{BASELINE} repeated once per batch.
For each trial, the explanation prompt is assembled by concatenating one labeled section per included module with fixed ordering and formatting; slice modules are rendered as original source lines annotated with absolute line numbers.

\paragraph{Slicing-based modules (SLICE\_*).}
Slice-based context modules provide failure-local code evidence by selecting subsets of source lines from the buggy module based on the triggering test.
We seed slicing by executing the test while tracing the buggy module and using the first exception line inside that module (fallback: last executed line).
\texttt{SLICE\_BLOCK} returns the innermost enclosing AST block containing the seed line.
\texttt{SLICE\_BACKWARD} approximates dynamic backward slicing by combining executed-line information with use/def tracking and a syntactic approximation of control dependence.
\texttt{SLICE\_FORWARD} is a scoped forward slice within the smallest enclosing function to capture potential downstream impact without uncontrolled growth.
\texttt{SLICE\_UNION} is the set union of backward and forward slice line sets.

\subsection{Models, Prompts, and Generation}
All LLM-mediated stages of the pipeline (explanation generation, LLM-as-a-judge scoring, and fix generation) use a single selected model per experimental sweep.
We evaluate three economically viable models: \texttt{gpt\allowbreak-5-mini} \citep{OpenRouter2026GPT5Mini}, \texttt{DeepSeek\allowbreak-V3.2} \citep{DeepSeekAI2026DeepSeekV32}, and \texttt{Grok\allowbreak-4.1-fast} \citep{OpenRouter2026Grok41Fast}.
To improve reproducibility, we use the lowest possible reasoning settings supported by each provider and set temperature $=0$ where configurable.
Nevertheless, strict determinism is not guaranteed due to service-level factors.

A \emph{trial} is one complete pipeline execution for a specific combination of defect, context configuration, model, and run identifier.
We execute three repeated runs per model with $\textit{run\_id}\in\{1,2,3\}$ under identical inputs.
To simplify automated processing and reduce ambiguity in downstream evaluation, we request structured outputs for explanations, judge scores, and fixes. Transient backend errors are handled via retry with exponential backoff.

\subsection{Evaluation Framework (Deterministic Metrics + LLM-as-a-Judge)}
We evaluate generated explanations with the hybrid framework introduced in Section~3 and operationalized here, scoring explanations on six binary criteria (C1--C6) and using both per-criterion rates and the total score \(q(e)\).
Criteria with direct signals are computed deterministically, while judgment-based criteria are evaluated with an \emph{LLM-as-a-judge} rubric \citep{Zheng2023MTBench,Liu2023GEval}.
Concretely, we compute readability (C1) using Flesch--Kincaid grade level and set $C1{=}1$ iff the grade level is $\le 12$ \citep{Kincaid1975FleschKincaid}; this threshold is a pragmatic readability heuristic for developer-facing prose rather than a universally validated cutoff for program explanations.
We compute contextual adequacy (C5) by counting explicit code references in the explanation (e.g., line references such as \texttt{L42} or function/method names) and set $C5{=}1$ iff the count is $\ge 2$, treating the threshold as a lightweight proxy for artifact grounding.
We then apply LLM-as-a-judge scoring for problem identification (C2), causal-chain clarity (C3), actionability (C4), and brevity (C6) under structured JSON outputs.

\subsection{Human Study for Judge Validation (RQ1)}
To validate whether LLM-as-a-judge labels approximate human judgment (RQ1), we conduct a user study with $12$ participants with a software engineering background (primarily graduate CS students, many with prior industry experience). All participants provided informed consent prior to the study.
The study objects are \emph{failure explanations}.
To limit participant workload, we evaluate a fixed subset of $8$ Python defects, sampled once from the $12$-defect dataset.
For each defect, participants label three independently generated \texttt{BASELINE} explanations produced under identical prompts and inputs (three runs of \texttt{gpt\allowbreak-5-mini} with $\textit{run\_id}=1\ldots 3$).
This yields $8 \times 3 = 24$ explanation items.
We then obtain LLM-as-a-judge labels for these same items by scoring them with each of the three judge models (\texttt{gpt\allowbreak-5-mini}, \texttt{DeepSeek\allowbreak-V3.2}, \texttt{Grok\allowbreak-4.1-fast}) under the same rubric, calling each judge once per participant-evaluation row, yielding $288$ judge calls per model ($12$ repeated labels per explanation item).
For each defect, participants are shown the defect identifier, a short reference root-cause description, and the three explanations displayed side-by-side.
These descriptions were derived during dataset curation from the original failure behavior, translated triggering test, and reference minimal fix.
Participants label each explanation as pass/fail on the four judgment-based criteria (C2/C3/C4/C6) and additionally rate labeling difficulty on a 5-point Likert scale.
To counterbalance order effects, the study randomizes defect order, the mapping from explanations to labels A/B/C, and the left-to-right column order. Randomization is participant-specific and deterministically seeded to support the resumption without changing assignments.

\subsection{Downstream Repair: Utility of Explanations}
To quantify downstream utility, we generate repairs in two conditions: (i) \texttt{NO\_\allowbreak{}EXPLANATION}, where the model receives the defect context only, and (ii) \texttt{WITH\_\allowbreak{}EXPLANATION}, where the generated explanation is appended to the same defect context.
Fix generation returns a function-only snippet, which we splice into the original module using AST location information so that only the target function body is replaced.
We functionally validate each candidate fix by executing the triggering test in isolation; an attempt passes if the test process exits with code $0$.
This validation checks whether the observed failure is removed, but it does not establish full regression correctness over the complete project test suite.

For passing fixes, we compare generated repairs against defect-specific reference minimal fixes using AST normalization to reduce superficial differences (e.g., local variable names) and compute diff-based similarity (line deviation and normalized Levenshtein distance~\cite{levenshtein1966binary}), change-localization measures, and Halstead delta-volume change~\cite{halstead1977elements}.
This comparison is a proxy for repair minimality and similarity to one accepted fix, not evidence that the reference is the only valid repair.
We relate repair outcomes to intermediate explanation quality by stratifying attempts by \(q(e)\) (and by individual criteria), testing whether higher explanation-score quartiles show higher repair success and more reference-like fixes, and whether low explanation-score quartiles can underperform relative to \texttt{NO\_\allowbreak{}EXPLANATION}.

%% file: 5_results/results.tex
\section{Results}\label{sec:results}

Unless stated otherwise, we emphasize effect sizes and uncertainty where available and use $p$-values as descriptive support.
For RQ3, we assess normality via Shapiro--Wilk and compare non-baseline two-way vs.\ three-way configuration-score distributions using a one-sided Welch's $t$-test (unequal variances); for RQ4, we report deltas with defect-bootstrap CIs.

\subsection{RQ1: Do Automated LLM-as-a-judge and Humans Produce Similar Evaluations of Failure Explanations?}
\textbf{Answer: Partially yes} -- LLM-as-a-judge is a reliable proxy for the causal and action-oriented criteria (C2--C4), but agreement drops for the more subjective brevity criterion (C6).
We validate the protocol by benchmarking judge labels against human ratings from the RQ1 user study ($n=24$ explanation items; $12$ participants; four criteria C2/C3/C4/C6) via four complementary agreement analyses (AQ1--AQ4).
AQ1 measures direct label agreement, AQ2 compares pass-rate prevalence, AQ3 checks repeated-rating consistency, and AQ4 captures perceived human labeling difficulty.

\paragraph{AQ1.}
\emph{Partially yes.} Agreement is high for the causal and action-oriented criteria (C2--C4, Acc $0.74$--$0.88$; Table~\ref{tab:rq1:llm_individual}), but only moderate for brevity (C6, Acc $\approx 0.58$).

\paragraph{AQ2.}
\emph{Yes.} All three judges are more pass-prone (i.e., more likely to assign a pass label) than humans, especially on C6 (LLM pass share $\approx 0.90$ vs.\ human pass share $0.58$; Table~\ref{tab:rq1:llm_individual_prev}).

\paragraph{AQ3.}
\emph{Yes.} LLM judges reach near-perfect unanimity on C2--C4 (up to $1.0$), whereas human unanimity is lower, particularly on C6 ($0.0$; Table~\ref{tab:rq1:llm_consistency}).

\paragraph{AQ4.}
\emph{Yes.} Brevity (C6) is rated hardest by humans (mean $2.82$; Table~\ref{tab:rq1:difficulty}), corroborating the lower agreement observed in AQ1--AQ3.

Across analyses, agreement is high for the causal and action-oriented criteria (C2--C4), while brevity (C6) shows lower agreement and higher human disagreement. LLM judges are also more pass-prone on C6, consistent with brevity being more subjective.
Overall, LLM-as-a-judge provides a scalable proxy for C2--C4, but subjective criteria such as C6 require calibration. \textbf{RQ1} results on the strong human--LLM agreement on causal and actionable criteria extend prior debugging studies on explanation usefulness~\cite{yan2024better,widyasari2024demystifying,salmon2025debugging}. This suggests that our criteria could allow LLM-as-a-judge to scale reliably as a human approximation.

\begin{table*}[tbp]
    \centering
    \caption{AQ1: Agreement between individual human labels and LLM-as-a-judge labels for C2--C4 and C6 ($n=288$ labels per criterion and model). Model columns report accuracy (Acc); ``LLM (avg.)'' is the macro-average across the three models.}
    \label{tab:rq1:llm_individual}
    \renewcommand{\arraystretch}{0.95}
    \footnotesize
    \setlength{\tabcolsep}{4pt}
    \begin{tabular*}{\textwidth}{@{}@{\extracolsep{\fill}}lcccc@{}}
        \toprule
        \textbf{Criterion} & \textbf{\texttt{gpt\allowbreak-5-mini}} & \textbf{\texttt{DeepSeek\allowbreak-V3.2}} & \textbf{\makecell{\texttt{Grok-}\\\texttt{4.1-fast}}} & \textbf{LLM (avg.)} \\
        \midrule
        \makecell[l]{\textbf{C2} \\ Problem identification} & 0.816 & 0.875 & 0.896 & 0.862 \\
        \makecell[l]{\textbf{C3} \\ Explanation clarity} & 0.708 & 0.753 & 0.767 & 0.743 \\
        \makecell[l]{\textbf{C4} \\ Actionability} & 0.847 & 0.875 & 0.903 & 0.875 \\
        \makecell[l]{\textbf{C6} \\ Brevity} & 0.587 & 0.573 & 0.576 & 0.579 \\
        \bottomrule
    \end{tabular*}
\end{table*}

\begin{table*}[tbp]
    \centering
    \caption{AQ2: Pass prevalence for individual human and LLM-as-a-judge labels for C2--C4 and C6 ($n=288$ labels per criterion and model). The human column reports the human pass share, model columns report LLM pass shares, and ``LLM (avg.)'' is the macro-average across the three models.}
    \label{tab:rq1:llm_individual_prev}
    \renewcommand{\arraystretch}{0.95}
    \footnotesize
    \setlength{\tabcolsep}{4pt}
    \begin{tabular*}{\textwidth}{@{}@{\extracolsep{\fill}}lccccc@{}}
        \toprule
        \textbf{Criterion} & \textbf{\makecell{Human\\pass share}} & \textbf{\texttt{gpt\allowbreak-5-mini}} & \textbf{\texttt{DeepSeek\allowbreak-V3.2}} & \textbf{\makecell{\texttt{Grok-}\\\texttt{4.1-fast}}} & \textbf{LLM (avg.)} \\
        \midrule
        \makecell[l]{\textbf{C2} \\ Problem identification} & 0.910 & 0.872 & 0.944 & 0.958 & 0.925 \\
        \makecell[l]{\textbf{C3} \\ Explanation clarity} & 0.781 & 0.878 & 0.944 & 0.979 & 0.934 \\
        \makecell[l]{\textbf{C4} \\ Actionability} & 0.903 & 0.903 & 0.965 & 1.000 & 0.956 \\
        \makecell[l]{\textbf{C6} \\ Brevity} & 0.583 & 0.837 & 0.878 & 0.993 & 0.903 \\
        \bottomrule
    \end{tabular*}
\end{table*}

\begin{table*}[tbp]
    \centering
    \caption{AQ3: Rating consistency for the same explanation item ($n=24$ items; 12 human votes and 12 repeated judge calls per model each). ``Unanim.'' is the fraction of unanimous items; ``Extrem.'' is $\mathrm{mean}(\max(p,1-p))$, where $p$ is the pass vote share.}
    \label{tab:rq1:llm_consistency}
    \renewcommand{\arraystretch}{0.95}
    \footnotesize
    \setlength{\tabcolsep}{4pt}
    \begin{tabular*}{\textwidth}{@{}@{\extracolsep{\fill}}lccccc@{}}
        \toprule
        \textbf{Criterion} & \textbf{Human} & \textbf{\texttt{gpt\allowbreak-5-mini}} & \textbf{\texttt{DeepSeek\allowbreak-V3.2}} & \textbf{\makecell{\texttt{Grok-}\\\texttt{4.1-fast}}} & \textbf{LLM (avg.)} \\
        \midrule
        \makecell[l]{\textbf{C2} \\ Problem identification} & \makecell{Unanim 0.417 \\ Extrem 0.910} & \makecell{Unanim 0.375 \\ Extrem 0.913} & \makecell{Unanim 0.917 \\ Extrem 0.979} & \makecell{Unanim 0.917 \\ Extrem 0.965} & \makecell{Unanim 0.736 \\ Extrem 0.953} \\
        \makecell[l]{\textbf{C3} \\ Explanation clarity} & \makecell{Unanim 0.167 \\ Extrem 0.781} & \makecell{Unanim 0.375 \\ Extrem 0.906} & \makecell{Unanim 0.917 \\ Extrem 0.979} & \makecell{Unanim 0.917 \\ Extrem 0.979} & \makecell{Unanim 0.736 \\ Extrem 0.955} \\
        \makecell[l]{\textbf{C4} \\ Actionability} & \makecell{Unanim 0.375 \\ Extrem 0.903} & \makecell{Unanim 0.375 \\ Extrem 0.910} & \makecell{Unanim 0.917 \\ Extrem 0.965} & \makecell{Unanim 1.000 \\ Extrem 1.000} & \makecell{Unanim 0.764 \\ Extrem 0.958} \\
        \makecell[l]{\textbf{C6} \\ Brevity} & \makecell{Unanim 0.000 \\ Extrem 0.667} & \makecell{Unanim 0.333 \\ Extrem 0.899} & \makecell{Unanim 0.833 \\ Extrem 0.962} & \makecell{Unanim 0.917 \\ Extrem 0.993} & \makecell{Unanim 0.694 \\ Extrem 0.951} \\
        \bottomrule
    \end{tabular*}
\end{table*}

\begin{table}[tbp]
    \centering
    \caption{AQ4: Per-criterion difficulty ratings from the user study (5-point Likert; 1 = easiest, 5 = hardest). Ratings are per participant and defect ($n=96$ participant$\times$defect cases). Values report the mean.}
    \label{tab:rq1:difficulty}
    \renewcommand{\arraystretch}{1.2}
    \small
    \setlength{\tabcolsep}{4pt}
    \begin{tabular*}{\linewidth}{@{}@{\extracolsep{\fill}}lc@{}}
        \toprule
        \textbf{Criterion} & \textbf{Mean difficulty} \\
        \midrule
        \makecell[l]{\textbf{C2} \\ Problem identification} & 2.24 \\
        \makecell[l]{\textbf{C3} \\ Explanation clarity} & 2.49 \\
        \makecell[l]{\textbf{C4} \\ Actionability} & 2.04 \\
        \makecell[l]{\textbf{C6} \\ Brevity} & 2.82 \\
        \bottomrule
    \end{tabular*}
\end{table}

\subsection{RQ2: Are Certain Evaluation Criteria More Discriminative Than Others?}
\textbf{Answer: Yes} -- We operationalize failure-\allowbreak{}explanation quality via six binary criteria (C1--C6). They cover presentation (C1 readability, C6 brevity), causal understanding (C2 problem identification, C3 explanation clarity), debugging support (C4 actionability), and grounding (C5 contextual adequacy).
To answer whether the criteria provide a discriminative signal, we link intermediate explanation quality to downstream repair success.
For each explanation we compute the total quality score \(q(e)=\sum_{i=1}^{6} C_i\) (range $0$--$6$) and stratify repair attempts into quartiles (Q1 lowest, Q4 highest) within each model and configuration batch (isolated, two-way, three-way).
Each batch pools attempts across all context configurations in that batch.
Table~\ref{tab:rq2:quartiles} shows that higher-scoring explanations consistently yield higher fix pass rates (Q4 $>$ Q1), while low-scoring explanations can underperform relative to \texttt{NO\_\allowbreak{}EXPLANATION} (e.g., \texttt{DeepSeek\allowbreak-V3.2} in three-way: Q1 $0.341$ vs.\ $0.500$), and high-scoring explanations outperform it (Q4 $0.891$ vs.\ $0.500$).
Across models, the Q4--Q1 differences are statistically significant in the composed settings (two-way/three-way; all $p<0.001$), but not in the isolated setting at $\alpha=0.05$ ($p=0.072$--$0.154$).
The observed pass-rate gaps are larger in the composed settings (18.0--55.0 percentage points) than in the isolated setting (11.1--13.6 percentage points).
Overall, the discrimination signal is strongest in the causal and action-oriented dimensions (C2--C4), while presentation and grounding criteria (C1/C6 and C5) are weaker indicators of downstream utility. The \textbf{RQ2} results extends previous work by showing that certain criteria are more discriminative for downstream repair success than presentation-focused criteria, reinforcing evidence that content-related dimensions drive usefulness~\cite{lu2025deepcrceval,yan2024better,widyasari2024demystifying,salmon2025debugging}.

\begin{table*}[tbp]
    \centering
    \caption{Downstream fix pass rates by explanation-quality quartile (Q1 vs.\ Q4) across models and context batches. Pass rates are shown as proportions with counts (passed/total). $p$ reports a two-sided two-proportion $z$-test comparing Q4 against Q1 (uncorrected); it indicates statistical significance rather than the size of the Q4--Q1 pass-rate gap. \texttt{NO\_\allowbreak{}EXPLANATION} is the no-explanation baseline.}
    \label{tab:rq2:quartiles}
    \renewcommand{\arraystretch}{1.2}
    \small
    \setlength{\tabcolsep}{4pt}
    \begin{tabular*}{\textwidth}{@{}@{\extracolsep{\fill}}llcrrrr@{}}
        \toprule
        \textbf{Model} &
        \textbf{\makecell[l]{Context\\batch}} &
        \textbf{Expl.?} &
        \textbf{\makecell[r]{Pass rate\\(passed/total)}} &
        \textbf{\makecell[r]{Q1 pass\\(passed/total)}} &
        \textbf{\makecell[r]{Q4 pass\\(passed/total)}} &
        \textbf{\makecell[r]{$p$\\(Q4 vs Q1)}} \\
        \midrule
        \multirow{4}{*}{\texttt{gpt\allowbreak-5-mini}} & \texttt{NO\_\allowbreak{}EXPLANATION} & No & 0.306 (11/36) & --- & --- & --- \\
        & isolated & Yes & 0.435 (141/324) & 0.296 (24/81) & 0.432 (35/81) & $0.072$ \\
        & two-way & Yes & 0.472 (493/1044) & 0.337 (88/261) & 0.517 (135/261) & $<0.001$ \\
        & three-way & Yes & 0.512 (1051/2052) & 0.352 (180/512) & 0.578 (297/514) & $<0.001$ \\
\midrule
        \multirow{4}{*}{\texttt{DeepSeek\allowbreak-V3.2}} & \texttt{NO\_\allowbreak{}EXPLANATION} & No & 0.500 (18/36) & --- & --- & --- \\
        & isolated & Yes & 0.448 (145/324) & 0.420 (34/81) & 0.543 (44/81) & $0.116$ \\
        & two-way & Yes & 0.500 (522/1044) & 0.360 (94/261) & 0.789 (206/261) & $<0.001$ \\
        & three-way & Yes & 0.553 (1134/2052) & 0.341 (175/513) & 0.891 (457/513) & $<0.001$ \\
        \midrule
        \multirow{4}{*}{\texttt{Grok\allowbreak-4.1-fast}} & \texttt{NO\_\allowbreak{}EXPLANATION} & No & 0.417 (15/36) & --- & --- & --- \\
        & isolated & Yes & 0.444 (144/324) & 0.383 (31/81) & 0.494 (40/81) & $0.154$ \\
        & two-way & Yes & 0.451 (471/1044) & 0.433 (113/261) & 0.628 (164/261) & $<0.001$ \\
        & three-way & Yes & 0.499 (1024/2052) & 0.376 (193/513) & 0.710 (364/513) & $<0.001$ \\
        \bottomrule
    \end{tabular*}
\end{table*}

\subsection{RQ3: Is the Quality of Failure Explanations Sensitive to Context Prompting?}
\textbf{Answer: Yes} -- We vary context configurations over eight modules (\texttt{CODE}, \texttt{ERROR}, \texttt{TEST}, \texttt{DOCSTRING}, and four slicing-based \texttt{SLICE\_*} modules) and score explanations on C1--C6.
We summarize quality as the expected total score \(E[\sum_{i=1}^6 C_i]\) (range $0$--$6$).
We compute per-defect expected scores as the sum of per-criterion pass rates (means of binary C1--C6) over replicated runs; configuration-level scores pooled across defects and runs.

\begin{figure*}[tbp]
    \centering
    \begin{minipage}[t]{0.32\textwidth}
        \centering
        \resizebox{0.95\linewidth}{!}{\input{figures/fig_rq3_isolated_total_boxplot_top3_gpt}}\\[-0.25em]
        {\scriptsize (a) \texttt{gpt\allowbreak-5-mini}}
    \end{minipage}\hfill%
    \begin{minipage}[t]{0.32\textwidth}
        \centering
        \resizebox{0.95\linewidth}{!}{\input{figures/fig_rq3_isolated_total_boxplot_top3_deepseek}}\\[-0.25em]
        {\scriptsize (b) \texttt{DeepSeek\allowbreak-V3.2}}
    \end{minipage}\hfill%
    \begin{minipage}[t]{0.32\textwidth}
        \centering
        \resizebox{0.95\linewidth}{!}{\input{figures/fig_rq3_isolated_total_boxplot_top3_grok}}\\[-0.25em]
        {\scriptsize (c) \texttt{Grok\allowbreak-4.1-fast}}
    \end{minipage}
    \Description{Three side-by-side boxplot panels showing the distribution of expected total explanation scores across defects for each model's top-3 isolated context modules and BASELINE.}
    \caption{Distributions of expected total explanation scores $E[\sum C_i]$ across defects for each model's top-3 isolated context modules (ranked by mean $E[\sum C_i]$) and \texttt{BASELINE}. Boxes show the IQR; center lines show the median; dots show the mean; whiskers show min/max. Abbreviations: \texttt{ERR}=\texttt{ERROR}, \texttt{DOC}=\texttt{DOCSTRING}, \texttt{BLK}=\texttt{SLICE\_BLOCK}, \texttt{BWD}=\texttt{SLICE\_BACKWARD}, \texttt{FWD}=\texttt{SLICE\_FORWARD}, \texttt{UNI}=\texttt{SLICE\_UNION}, \texttt{BASE}=\texttt{BASELINE}.}
    \label{fig:rq3:isolated_total}
\end{figure*}
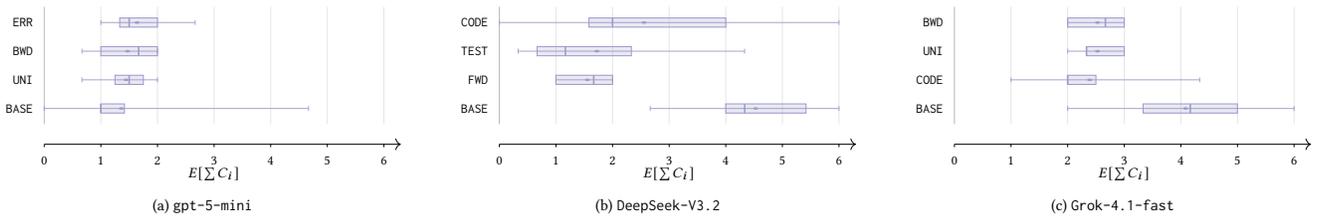

Figure~\ref{fig:rq3:isolated_total} shows score distributions for each model's top-3 isolated context modules (ranked by mean $E[\sum C_i]$) and \texttt{BASELINE} (all modules); the full isolated set is omitted for space.

In isolated settings (Figure~\ref{fig:rq3:isolated_total}), \texttt{BASELINE} (all modules) yields the highest quality for \texttt{DeepSeek\allowbreak-V3.2} ($4.528$) and \texttt{Grok\allowbreak-4.1-fast} ($4.082$).
\texttt{gpt\allowbreak-5-mini} performs best with the atomic \texttt{ERROR} module ($1.640$) and remains low even under \texttt{BASELINE} ($1.361$).
Slice modules achieve their isolated scores primarily by improving grounding (C5).

\begin{figure*}[tbp]
    \centering
    \input{figures/fig_rq3_config_score_dist_boxplot}
    \Description{Three side-by-side boxplot panels of configuration-level expected total explanation scores for gpt-5-mini, DeepSeek-V3.2, and Grok-4.1-fast in isolated, two-way, and three-way batches. Open circles mark the BASELINE score for each batch.}
    \caption{Distributions of configuration-level expected total explanation scores $E[\sum C_i]$ in isolated, two-way, and three-way batches. Open circles mark the \texttt{BASELINE} score for each batch. Boxes show the IQR; center lines show the median; dots show the mean; whiskers show min/max.}
    \label{fig:rq3:config_score_dist_boxplot}
\end{figure*}
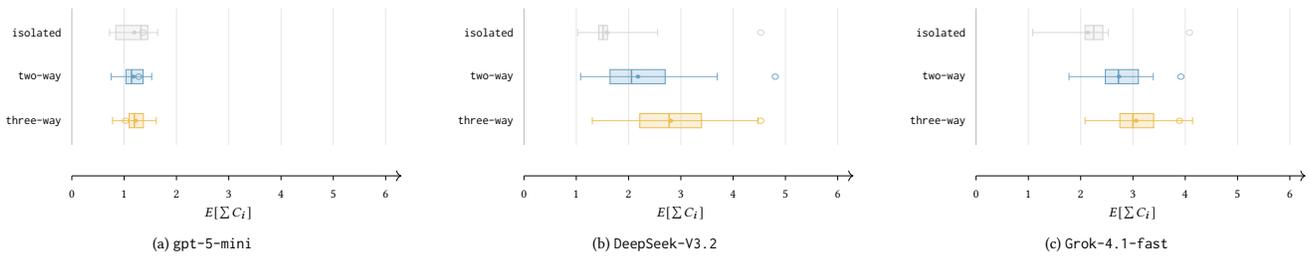

Moving beyond isolated settings, Figure~\ref{fig:rq3:config_score_dist_boxplot} summarizes the distribution of configuration-level expected total explanation scores across the isolated, two-way, and three-way batches.
For \texttt{DeepSeek\allowbreak-V3.2} and \texttt{Grok\allowbreak-4.1-fast}, composed batches shift upward relative to isolated, with three-way typically exceeding two-way.
In contrast, \texttt{gpt\allowbreak-5-mini} remains low across batches and shows only limited gains from composition.
Because the isolated batch contains only eight non-baseline configurations ($n=8$), we treat the isolated distributions in Figure~\ref{fig:rq3:config_score_dist_boxplot} descriptively and do not conduct hypothesis tests for that condition.
We restrict inference to the two-way vs.\ three-way comparison and test this contrast separately per model. Since this is one planned test per model ($m=1$), we report uncorrected $p$-values.
A Shapiro--Wilk test does not indicate significant departures from normality for any of the six non-baseline two-way and three-way configuration-score distributions (all $p>0.2$).
Across non-baseline configuration-level scores (two-way: $n=28$ configurations; three-way: $n=56$), we compare two-way vs.\ three-way distributions per model using the unequal-variance one-sided Welch's $t$-test ($H_1:\mu_{\text{two-way}} < \mu_{\text{three-way}}$). This confirms higher mean scores in three-way than two-way batches for \texttt{DeepSeek\allowbreak-V3.2} ($p=2.9\times 10^{-4}$) and \texttt{Grok\allowbreak-4.1-fast} ($p=8.1\times 10^{-4}$), but not for \texttt{gpt\allowbreak-5-mini} ($p=0.165$). These conclusions are unchanged under a Bonferroni correction across the three model-level tests.

To interpret composed settings, we report marginal effects $\Delta$ of including each atomic module in two-way and three-way compositions; positive values indicate higher-scoring explanations when the module is included.
For \texttt{DeepSeek\allowbreak-V3.2}, the largest positive effects come from adding \texttt{CODE} and \texttt{TEST} ($\Delta \approx +1.18$ and $+0.97$ in two-way; $+0.99$ and $+0.80$ in three-way), while \texttt{DOCSTRING} consistently has negative marginal effects ($\Delta \approx -0.44$ to $-0.61$).
For \texttt{Grok\allowbreak-4.1-fast}, \texttt{CODE}/\texttt{TEST} remain beneficial and \texttt{SLICE\_UNION} is the strongest slice factor in two-way compositions ($\Delta \approx +0.47$).
In contrast, \texttt{gpt\allowbreak-5-mini} benefits mainly from slices (especially \texttt{SLICE\_UNION}). \textbf{RQ3} results add confirmatory evidence on the sensitivity of LLM output to causal context  manipulation~\cite{Holloway2024ContextGranularityAPR,guan2025the,rando2025longcodebench}.

\subsection{RQ4: Do Higher Explanation-Quality Quartiles Show Generated Fixes That Better Match a Reference Minimal Fix?}
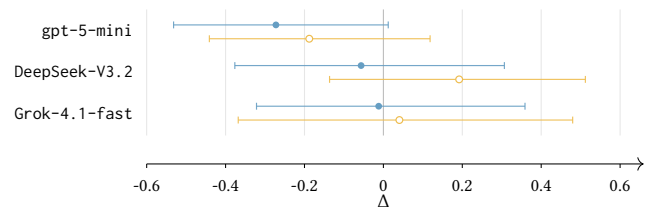
\begin{figure}[tbp]
    \centering
    \input{figures/fig_rq4_delta_minimality}
    \caption{Minimality effect sizes ($\Delta =$ Q4--Q1) for passing fixes in composed batches. Blue filled circles denote two-way batches; orange open circles denote three-way batches. Whiskers show Bonferroni-adjusted defect-bootstrap CIs ($m=2$ per model). Positive values indicate that higher-quality explanations coincide with a higher minimal-fix rate.}
    \Description{Effect-size plot (Q4--Q1) of minimal-fix rate by model for two-way (blue filled circles) and three-way (orange open circles) composed batches, with whiskers showing Bonferroni-adjusted defect-bootstrap confidence intervals.}
    \label{fig:rq4:minimality_delta}
\end{figure}

\textbf{Answer: Partially yes} -- We relate explanation quality to fix minimality and similarity to the defect-specific reference minimal fix.
For each model and configuration batch, we sort scored attempts by total explanation score \(q(e)\), split them into quartiles, and report metrics over \emph{passing} fixes (Q1 vs.\ Q4).
Figure~\ref{fig:rq4:minimality_delta} summarizes minimality effect sizes ($\Delta =$ Q4--Q1) for the composed batches.
We omit isolated because it contains only eight configurations and yields noisier deltas. Subsequent comparisons focus on two-way vs.\ three-way.

Results are model-dependent, and evidence for a clear minimality effect is limited: across all models and composed batches, the Bonferroni-adjusted defect-bootstrap CIs in Figure~\ref{fig:rq4:minimality_delta} overlap $0$.
Point estimates improve for \texttt{DeepSeek\allowbreak-V3.2} (e.g., $0.238 \to 0.430$ in three-way), change only slightly for \texttt{Grok\allowbreak-4.1-fast} (e.g., $0.443 \to 0.483$), and trend downward for \texttt{gpt\allowbreak-5-mini} (e.g., $0.379 \to 0.191$), but all remain uncertain.

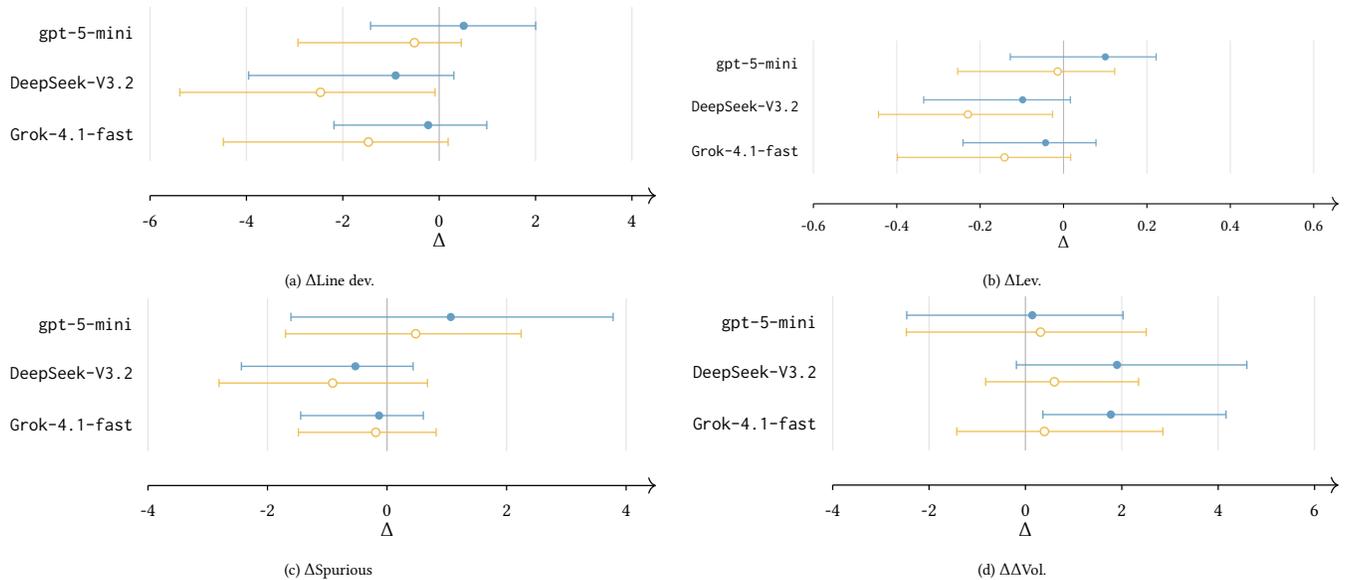
\begin{figure*}[tbp]
    \centering
    \begin{minipage}[t]{0.49\textwidth}
        \centering
        \input{figures/fig_rq4_delta_line_deviation}
        \vspace{-0.4em}
        {\scriptsize (a) $\Delta$Line dev.}\par
    \end{minipage}\hfill%
    \begin{minipage}[t]{0.49\textwidth}
        \centering
        \input{figures/fig_rq4_delta_levenshtein}
        \vspace{-0.4em}
        {\scriptsize (b) $\Delta$Lev.}\par
    \end{minipage}

    \vspace{0.6em}

    \begin{minipage}[t]{0.49\textwidth}
        \centering
        \input{figures/fig_rq4_delta_spurious}
        \vspace{-0.4em}
        {\scriptsize (c) $\Delta$Spurious}\par
    \end{minipage}\hfill%
    \begin{minipage}[t]{0.49\textwidth}
        \centering
        \input{figures/fig_rq4_delta_delta_volume}
        \vspace{-0.4em}
        {\scriptsize (d) $\Delta\Delta$Vol.}\par
    \end{minipage}
    \caption{Effect sizes comparing high- vs.\ low-quality explanations on passing fixes in composed batches ($\Delta =$ Q4--Q1). Blue filled circles denote two-way batches; orange open circles denote three-way batches. Whiskers show Bonferroni-adjusted defect-bootstrap CIs ($m=2$ per model). Negative values for line deviation, normalized Levenshtein, and spurious changed lines indicate more reference-like fixes; positive $\Delta\Delta$Vol.\ indicates a larger complexity increase (Fix $-$ Baseline).}
    \Description{Four-panel effect-size plot (Q4--Q1) of fix similarity (line deviation and normalized Levenshtein), spurious changed lines, and complexity increase (Halstead delta volume) by model. Two-way batches are shown with blue filled circles; three-way batches are shown with orange open circles. Whiskers show Bonferroni-adjusted defect-bootstrap confidence intervals.}
    \label{fig:rq4:similarity_deltas}
\end{figure*}

Beyond exact minimality, higher-quality explanations can coincide with smaller diffs to the reference fix and fewer spurious edits for \texttt{DeepSeek\allowbreak-V3.2} and \texttt{Grok\allowbreak-4.1-fast} (Figure~\ref{fig:rq4:similarity_deltas}), while \texttt{gpt\allowbreak-5-mini} shows more spurious changes.
Figure~\ref{fig:rq4:similarity_deltas} reports Q4--Q1 effect sizes for diff-based similarity ($\Delta$Line dev., $\Delta$Lev.), change localization ($\Delta$Spurious), and complexity ($\Delta\Delta$Vol.) on passing fixes in composed batches.
Whiskers show Bonferroni-adjusted defect-bootstrap CIs across the two composed batches per model ($m=2$); we conservatively highlight effects whose CI excludes $0$.
Clear similarity improvements are limited to \texttt{DeepSeek\allowbreak-V3.2} in the three-way batch, where CIs are below $0$ for $\Delta$Line dev.\ and $\Delta$Lev.; $\Delta$Spurious remains inconclusive for all models, and the only complexity increase with a CI above $0$ occurs for \texttt{Grok\allowbreak-4.1-fast} in two-way.
Overall, similarity/noise effects are model- and batch-dependent and often uncertain. The \textbf{RQ4} results partially corroborate prior evidence that explanations can improve fix accuracy~\cite{adriano2022microtasking}, but add that quality gains translate into more reference-like fixes only conditionally across models and context batches.

%% file: figures/fig_rq3_isolated_total_boxplot_top3_gpt.tex
\begin{tikzpicture}[x=0.8888888888888887cm, y=0.45cm, font=\scriptsize]
    \definecolor{stroke2}{RGB}{230,159,0}
    \definecolor{stroke3}{RGB}{117,112,179}
    \draw[gray!55, solid] (0.000,0.550) -- (0.000,-3.550);
    \draw (0.000,-4.250) -- (0.000,-4.350);
    \node[anchor=north] at (0.000,-4.430) {0};
    \draw[gray!20, solid] (1.000,0.550) -- (1.000,-3.550);
    \draw (1.000,-4.250) -- (1.000,-4.350);
    \node[anchor=north] at (1.000,-4.430) {1};
    \draw[gray!20, solid] (2.000,0.550) -- (2.000,-3.550);
    \draw (2.000,-4.250) -- (2.000,-4.350);
    \node[anchor=north] at (2.000,-4.430) {2};
    \draw[gray!20, solid] (3.000,0.550) -- (3.000,-3.550);
    \draw (3.000,-4.250) -- (3.000,-4.350);
    \node[anchor=north] at (3.000,-4.430) {3};
    \draw[gray!20, solid] (4.000,0.550) -- (4.000,-3.550);
    \draw (4.000,-4.250) -- (4.000,-4.350);
    \node[anchor=north] at (4.000,-4.430) {4};
    \draw[gray!20, solid] (5.000,0.550) -- (5.000,-3.550);
    \draw (5.000,-4.250) -- (5.000,-4.350);
    \node[anchor=north] at (5.000,-4.430) {5};
    \draw[gray!20, solid] (6.000,0.550) -- (6.000,-3.550);
    \draw (6.000,-4.250) -- (6.000,-4.350);
    \node[anchor=north] at (6.000,-4.430) {6};
    \draw[->] (0.000,-4.250) -- (6.300,-4.250);
    \node[anchor=north] at (3.000,-4.800) {$E[\sum C_i]$};

    \def\boxHalf{0.160}\def\capHalf{0.085}\def\dotR{0.045}\def\baseR{0.060}
    \def\whiskLW{0.45pt}\def\whiskOp{0.90}\def\boxLW{0.45pt}\def\medLW{0.70pt}\def\baseLW{0.45pt}\def\boxOp{0.22}

    \node[anchor=east] at (-0.072,-0.000) {\texttt{ERR}};
    \draw[line width=\whiskLW, opacity=\whiskOp, draw=stroke3!70] (1.00000,-0.000) -- (2.66500,-0.000);
    \draw[line width=\whiskLW, opacity=\whiskOp, draw=stroke3!70] (1.00000,-0.085) -- (1.00000,0.085);
    \draw[line width=\whiskLW, opacity=\whiskOp, draw=stroke3!70] (2.66500,-0.085) -- (2.66500,0.085);
    \path[fill=stroke3!70, opacity=\boxOp] (1.33300,-0.160) rectangle (2.00000,0.160);
    \draw[line width=\boxLW, draw=stroke3!70] (1.33300,-0.160) rectangle (2.00000,0.160);
    \draw[line width=\medLW, draw=stroke3!70] (1.50000,-0.160) -- (1.50000,0.160);
    \fill[stroke3!70] (1.63867,-0.000) circle (\dotR);

    \node[anchor=east] at (-0.072,-1.000) {\texttt{BWD}};
    \draw[line width=\whiskLW, opacity=\whiskOp, draw=stroke3!70] (0.66600,-1.000) -- (2.00000,-1.000);
    \draw[line width=\whiskLW, opacity=\whiskOp, draw=stroke3!70] (0.66600,-1.085) -- (0.66600,-0.915);
    \draw[line width=\whiskLW, opacity=\whiskOp, draw=stroke3!70] (2.00000,-1.085) -- (2.00000,-0.915);
    \path[fill=stroke3!70, opacity=\boxOp] (1.00000,-1.160) rectangle (2.00000,-0.840);
    \draw[line width=\boxLW, draw=stroke3!70] (1.00000,-1.160) rectangle (2.00000,-0.840);
    \draw[line width=\medLW, draw=stroke3!70] (1.66700,-1.160) -- (1.66700,-0.840);
    \fill[stroke3!70] (1.47225,-1.000) circle (\dotR);

    \node[anchor=east] at (-0.072,-2.000) {\texttt{UNI}};
    \draw[line width=\whiskLW, opacity=\whiskOp, draw=stroke3!70] (0.66600,-2.000) -- (2.00000,-2.000);
    \draw[line width=\whiskLW, opacity=\whiskOp, draw=stroke3!70] (0.66600,-2.085) -- (0.66600,-1.915);
    \draw[line width=\whiskLW, opacity=\whiskOp, draw=stroke3!70] (2.00000,-2.085) -- (2.00000,-1.915);
    \path[fill=stroke3!70, opacity=\boxOp] (1.24975,-2.160) rectangle (1.75025,-1.840);
    \draw[line width=\boxLW, draw=stroke3!70] (1.24975,-2.160) rectangle (1.75025,-1.840);
    \draw[line width=\medLW, draw=stroke3!70] (1.50050,-2.160) -- (1.50050,-1.840);
    \fill[stroke3!70] (1.44450,-2.000) circle (\dotR);

    \node[anchor=east] at (-0.072,-3.000) {\texttt{BASE}};
    \draw[line width=\whiskLW, opacity=\whiskOp, draw=stroke3!70] (0.00000,-3.000) -- (4.66800,-3.000);
    \draw[line width=\whiskLW, opacity=\whiskOp, draw=stroke3!70] (0.00000,-3.085) -- (0.00000,-2.915);
    \draw[line width=\whiskLW, opacity=\whiskOp, draw=stroke3!70] (4.66800,-3.085) -- (4.66800,-2.915);
    \path[fill=stroke3!70, opacity=\boxOp] (1.00000,-3.160) rectangle (1.41650,-2.840);
    \draw[line width=\boxLW, draw=stroke3!70] (1.00000,-3.160) rectangle (1.41650,-2.840);
    \draw[line width=\medLW, draw=stroke3!70] (1.00000,-3.160) -- (1.00000,-2.840);
    \fill[stroke3!70] (1.36108,-3.000) circle (\dotR);

\end{tikzpicture}

%% file: figures/fig_rq3_isolated_total_boxplot_top3_deepseek.tex
\begin{tikzpicture}[x=0.8888888888888887cm, y=0.45cm, font=\scriptsize]
    \definecolor{stroke2}{RGB}{230,159,0}
    \definecolor{stroke3}{RGB}{117,112,179}
    \draw[gray!55, solid] (0.000,0.550) -- (0.000,-3.550);
    \draw (0.000,-4.250) -- (0.000,-4.350);
    \node[anchor=north] at (0.000,-4.430) {0};
    \draw[gray!20, solid] (1.000,0.550) -- (1.000,-3.550);
    \draw (1.000,-4.250) -- (1.000,-4.350);
    \node[anchor=north] at (1.000,-4.430) {1};
    \draw[gray!20, solid] (2.000,0.550) -- (2.000,-3.550);
    \draw (2.000,-4.250) -- (2.000,-4.350);
    \node[anchor=north] at (2.000,-4.430) {2};
    \draw[gray!20, solid] (3.000,0.550) -- (3.000,-3.550);
    \draw (3.000,-4.250) -- (3.000,-4.350);
    \node[anchor=north] at (3.000,-4.430) {3};
    \draw[gray!20, solid] (4.000,0.550) -- (4.000,-3.550);
    \draw (4.000,-4.250) -- (4.000,-4.350);
    \node[anchor=north] at (4.000,-4.430) {4};
    \draw[gray!20, solid] (5.000,0.550) -- (5.000,-3.550);
    \draw (5.000,-4.250) -- (5.000,-4.350);
    \node[anchor=north] at (5.000,-4.430) {5};
    \draw[gray!20, solid] (6.000,0.550) -- (6.000,-3.550);
    \draw (6.000,-4.250) -- (6.000,-4.350);
    \node[anchor=north] at (6.000,-4.430) {6};
    \draw[->] (0.000,-4.250) -- (6.300,-4.250);
    \node[anchor=north] at (3.000,-4.800) {$E[\sum C_i]$};

    \def\boxHalf{0.160}\def\capHalf{0.085}\def\dotR{0.045}\def\baseR{0.060}
    \def\whiskLW{0.45pt}\def\whiskOp{0.90}\def\boxLW{0.45pt}\def\medLW{0.70pt}\def\baseLW{0.45pt}\def\boxOp{0.22}

    \node[anchor=east] at (-0.072,-0.000) {\texttt{CODE}};
    \draw[line width=\whiskLW, opacity=\whiskOp, draw=stroke3!70] (0.00000,-0.000) -- (6.00000,-0.000);
    \draw[line width=\whiskLW, opacity=\whiskOp, draw=stroke3!70] (0.00000,-0.085) -- (0.00000,0.085);
    \draw[line width=\whiskLW, opacity=\whiskOp, draw=stroke3!70] (6.00000,-0.085) -- (6.00000,0.085);
    \path[fill=stroke3!70, opacity=\boxOp] (1.58350,-0.160) rectangle (4.00100,0.160);
    \draw[line width=\boxLW, draw=stroke3!70] (1.58350,-0.160) rectangle (4.00100,0.160);
    \draw[line width=\medLW, draw=stroke3!70] (2.00000,-0.160) -- (2.00000,0.160);
    \fill[stroke3!70] (2.55575,-0.000) circle (\dotR);

    \node[anchor=east] at (-0.072,-1.000) {\texttt{TEST}};
    \draw[line width=\whiskLW, opacity=\whiskOp, draw=stroke3!70] (0.33300,-1.000) -- (4.33300,-1.000);
    \draw[line width=\whiskLW, opacity=\whiskOp, draw=stroke3!70] (0.33300,-1.085) -- (0.33300,-0.915);
    \draw[line width=\whiskLW, opacity=\whiskOp, draw=stroke3!70] (4.33300,-1.085) -- (4.33300,-0.915);
    \path[fill=stroke3!70, opacity=\boxOp] (0.66700,-1.160) rectangle (2.33225,-0.840);
    \draw[line width=\boxLW, draw=stroke3!70] (0.66700,-1.160) rectangle (2.33225,-0.840);
    \draw[line width=\medLW, draw=stroke3!70] (1.16650,-1.160) -- (1.16650,-0.840);
    \fill[stroke3!70] (1.72200,-1.000) circle (\dotR);

    \node[anchor=east] at (-0.072,-2.000) {\texttt{FWD}};
    \draw[line width=\whiskLW, opacity=\whiskOp, draw=stroke3!70] (1.00000,-2.000) -- (2.00000,-2.000);
    \draw[line width=\whiskLW, opacity=\whiskOp, draw=stroke3!70] (1.00000,-2.085) -- (1.00000,-1.915);
    \draw[line width=\whiskLW, opacity=\whiskOp, draw=stroke3!70] (2.00000,-2.085) -- (2.00000,-1.915);
    \path[fill=stroke3!70, opacity=\boxOp] (1.00000,-2.160) rectangle (2.00000,-1.840);
    \draw[line width=\boxLW, draw=stroke3!70] (1.00000,-2.160) rectangle (2.00000,-1.840);
    \draw[line width=\medLW, draw=stroke3!70] (1.66700,-2.160) -- (1.66700,-1.840);
    \fill[stroke3!70] (1.55567,-2.000) circle (\dotR);

    \node[anchor=east] at (-0.072,-3.000) {\texttt{BASE}};
    \draw[line width=\whiskLW, opacity=\whiskOp, draw=stroke3!70] (2.66500,-3.000) -- (6.00000,-3.000);
    \draw[line width=\whiskLW, opacity=\whiskOp, draw=stroke3!70] (2.66500,-3.085) -- (2.66500,-2.915);
    \draw[line width=\whiskLW, opacity=\whiskOp, draw=stroke3!70] (6.00000,-3.085) -- (6.00000,-2.915);
    \path[fill=stroke3!70, opacity=\boxOp] (4.00000,-3.160) rectangle (5.41725,-2.840);
    \draw[line width=\boxLW, draw=stroke3!70] (4.00000,-3.160) rectangle (5.41725,-2.840);
    \draw[line width=\medLW, draw=stroke3!70] (4.33400,-3.160) -- (4.33400,-2.840);
    \fill[stroke3!70] (4.52792,-3.000) circle (\dotR);

\end{tikzpicture}

%% file: figures/fig_rq3_isolated_total_boxplot_top3_grok.tex
\begin{tikzpicture}[x=0.8888888888888887cm, y=0.45cm, font=\scriptsize]
    \definecolor{stroke2}{RGB}{230,159,0}
    \definecolor{stroke3}{RGB}{117,112,179}
    \draw[gray!55, solid] (0.000,0.550) -- (0.000,-3.550);
    \draw (0.000,-4.250) -- (0.000,-4.350);
    \node[anchor=north] at (0.000,-4.430) {0};
    \draw[gray!20, solid] (1.000,0.550) -- (1.000,-3.550);
    \draw (1.000,-4.250) -- (1.000,-4.350);
    \node[anchor=north] at (1.000,-4.430) {1};
    \draw[gray!20, solid] (2.000,0.550) -- (2.000,-3.550);
    \draw (2.000,-4.250) -- (2.000,-4.350);
    \node[anchor=north] at (2.000,-4.430) {2};
    \draw[gray!20, solid] (3.000,0.550) -- (3.000,-3.550);
    \draw (3.000,-4.250) -- (3.000,-4.350);
    \node[anchor=north] at (3.000,-4.430) {3};
    \draw[gray!20, solid] (4.000,0.550) -- (4.000,-3.550);
    \draw (4.000,-4.250) -- (4.000,-4.350);
    \node[anchor=north] at (4.000,-4.430) {4};
    \draw[gray!20, solid] (5.000,0.550) -- (5.000,-3.550);
    \draw (5.000,-4.250) -- (5.000,-4.350);
    \node[anchor=north] at (5.000,-4.430) {5};
    \draw[gray!20, solid] (6.000,0.550) -- (6.000,-3.550);
    \draw (6.000,-4.250) -- (6.000,-4.350);
    \node[anchor=north] at (6.000,-4.430) {6};
    \draw[->] (0.000,-4.250) -- (6.300,-4.250);
    \node[anchor=north] at (3.000,-4.800) {$E[\sum C_i]$};

    \def\boxHalf{0.160}\def\capHalf{0.085}\def\dotR{0.045}\def\baseR{0.060}
    \def\whiskLW{0.45pt}\def\whiskOp{0.90}\def\boxLW{0.45pt}\def\medLW{0.70pt}\def\baseLW{0.45pt}\def\boxOp{0.22}

    \node[anchor=east] at (-0.072,-0.000) {\texttt{BWD}};
    \draw[line width=\whiskLW, opacity=\whiskOp, draw=stroke3!70] (2.00000,-0.000) -- (3.00000,-0.000);
    \draw[line width=\whiskLW, opacity=\whiskOp, draw=stroke3!70] (2.00000,-0.085) -- (2.00000,0.085);
    \draw[line width=\whiskLW, opacity=\whiskOp, draw=stroke3!70] (3.00000,-0.085) -- (3.00000,0.085);
    \path[fill=stroke3!70, opacity=\boxOp] (2.00000,-0.160) rectangle (3.00000,0.160);
    \draw[line width=\boxLW, draw=stroke3!70] (2.00000,-0.160) rectangle (3.00000,0.160);
    \draw[line width=\medLW, draw=stroke3!70] (2.66700,-0.160) -- (2.66700,0.160);
    \fill[stroke3!70] (2.52783,-0.000) circle (\dotR);

    \node[anchor=east] at (-0.072,-1.000) {\texttt{UNI}};
    \draw[line width=\whiskLW, opacity=\whiskOp, draw=stroke3!70] (2.00000,-1.000) -- (3.00000,-1.000);
    \draw[line width=\whiskLW, opacity=\whiskOp, draw=stroke3!70] (2.00000,-1.085) -- (2.00000,-0.915);
    \draw[line width=\whiskLW, opacity=\whiskOp, draw=stroke3!70] (3.00000,-1.085) -- (3.00000,-0.915);
    \path[fill=stroke3!70, opacity=\boxOp] (2.33300,-1.160) rectangle (3.00000,-0.840);
    \draw[line width=\boxLW, draw=stroke3!70] (2.33300,-1.160) rectangle (3.00000,-0.840);
    \draw[line width=\medLW, draw=stroke3!70] (2.33300,-1.160) -- (2.33300,-0.840);
    \fill[stroke3!70] (2.52767,-1.000) circle (\dotR);

    \node[anchor=east] at (-0.072,-2.000) {\texttt{CODE}};
    \draw[line width=\whiskLW, opacity=\whiskOp, draw=stroke3!70] (1.00000,-2.000) -- (4.33300,-2.000);
    \draw[line width=\whiskLW, opacity=\whiskOp, draw=stroke3!70] (1.00000,-2.085) -- (1.00000,-1.915);
    \draw[line width=\whiskLW, opacity=\whiskOp, draw=stroke3!70] (4.33300,-2.085) -- (4.33300,-1.915);
    \path[fill=stroke3!70, opacity=\boxOp] (2.00000,-2.160) rectangle (2.49975,-1.840);
    \draw[line width=\boxLW, draw=stroke3!70] (2.00000,-2.160) rectangle (2.49975,-1.840);
    \draw[line width=\medLW, draw=stroke3!70] (2.00000,-2.160) -- (2.00000,-1.840);
    \fill[stroke3!70] (2.38875,-2.000) circle (\dotR);

    \node[anchor=east] at (-0.072,-3.000) {\texttt{BASE}};
    \draw[line width=\whiskLW, opacity=\whiskOp, draw=stroke3!70] (2.00000,-3.000) -- (6.00000,-3.000);
    \draw[line width=\whiskLW, opacity=\whiskOp, draw=stroke3!70] (2.00000,-3.085) -- (2.00000,-2.915);
    \draw[line width=\whiskLW, opacity=\whiskOp, draw=stroke3!70] (6.00000,-3.085) -- (6.00000,-2.915);
    \path[fill=stroke3!70, opacity=\boxOp] (3.33300,-3.160) rectangle (5.00000,-2.840);
    \draw[line width=\boxLW, draw=stroke3!70] (3.33300,-3.160) rectangle (5.00000,-2.840);
    \draw[line width=\medLW, draw=stroke3!70] (4.16650,-3.160) -- (4.16650,-2.840);
    \fill[stroke3!70] (4.08317,-3.000) circle (\dotR);

\end{tikzpicture}

%% file: figures/fig_rq3_config_score_dist_boxplot.tex
\centering
\begin{minipage}[t]{0.32\textwidth}
    \centering
    \resizebox{0.95\linewidth}{!}{\input{figures/fig_rq3_config_score_dist_boxplot_gpt}}\\[-0.25em]
    {\scriptsize (a) \texttt{gpt\allowbreak-5-mini}}
\end{minipage}\hfill%
\begin{minipage}[t]{0.32\textwidth}
    \centering
    \resizebox{0.95\linewidth}{!}{\input{figures/fig_rq3_config_score_dist_boxplot_deepseek}}\\[-0.25em]
    {\scriptsize (b) \texttt{DeepSeek\allowbreak-V3.2}}
\end{minipage}\hfill%
\begin{minipage}[t]{0.32\textwidth}
    \centering
    \resizebox{0.95\linewidth}{!}{\input{figures/fig_rq3_config_score_dist_boxplot_grok}}\\[-0.25em]
    {\scriptsize (c) \texttt{Grok\allowbreak-4.1-fast}}
\end{minipage}

%% file: figures/fig_rq3_config_score_dist_boxplot_gpt.tex
\begin{tikzpicture}[x=0.8888888888888887cm, y=0.75cm, font=\scriptsize]
    \definecolor{stroke2}{RGB}{230,159,0}
    \definecolor{stroke3}{RGB}{117,112,179}
    \draw[gray!55, solid] (0.000,0.550) -- (0.000,-2.550);
    \draw (0.000,-3.250) -- (0.000,-3.350);
    \node[anchor=north] at (0.000,-3.430) {0};
    \draw[gray!20, solid] (1.000,0.550) -- (1.000,-2.550);
    \draw (1.000,-3.250) -- (1.000,-3.350);
    \node[anchor=north] at (1.000,-3.430) {1};
    \draw[gray!20, solid] (2.000,0.550) -- (2.000,-2.550);
    \draw (2.000,-3.250) -- (2.000,-3.350);
    \node[anchor=north] at (2.000,-3.430) {2};
    \draw[gray!20, solid] (3.000,0.550) -- (3.000,-2.550);
    \draw (3.000,-3.250) -- (3.000,-3.350);
    \node[anchor=north] at (3.000,-3.430) {3};
    \draw[gray!20, solid] (4.000,0.550) -- (4.000,-2.550);
    \draw (4.000,-3.250) -- (4.000,-3.350);
    \node[anchor=north] at (4.000,-3.430) {4};
    \draw[gray!20, solid] (5.000,0.550) -- (5.000,-2.550);
    \draw (5.000,-3.250) -- (5.000,-3.350);
    \node[anchor=north] at (5.000,-3.430) {5};
    \draw[gray!20, solid] (6.000,0.550) -- (6.000,-2.550);
    \draw (6.000,-3.250) -- (6.000,-3.350);
    \node[anchor=north] at (6.000,-3.430) {6};
    \draw[->] (0.000,-3.250) -- (6.300,-3.250);
    \node[anchor=north] at (3.000,-3.800) {$E[\sum C_i]$};

    \def\boxHalf{0.160}\def\capHalf{0.085}\def\dotR{0.045}\def\baseR{0.060}
    \def\whiskLW{0.45pt}\def\whiskOp{0.90}\def\boxLW{0.45pt}\def\medLW{0.70pt}\def\baseLW{0.45pt}\def\boxOp{0.22}

    \node[anchor=east] at (-0.072,-0.000) {\texttt{isolated}};
    \draw[line width=\whiskLW, opacity=\whiskOp, draw=gray!35] (0.72200,-0.000) -- (1.64000,-0.000);
    \draw[line width=\whiskLW, opacity=\whiskOp, draw=gray!35] (0.72200,-0.085) -- (0.72200,0.085);
    \draw[line width=\whiskLW, opacity=\whiskOp, draw=gray!35] (1.64000,-0.085) -- (1.64000,0.085);
    \path[fill=gray!35, opacity=\boxOp] (0.84025,-0.160) rectangle (1.45100,0.160);
    \draw[line width=\boxLW, draw=gray!35] (0.84025,-0.160) rectangle (1.45100,0.160);
    \draw[line width=\medLW, draw=gray!35] (1.31950,-0.160) -- (1.31950,0.160);
    \fill[gray!35] (1.19450,-0.000) circle (\dotR);
    \draw[line width=\baseLW, opacity=\whiskOp, draw=gray!35] (1.36100,-0.000) circle (\baseR);

    \node[anchor=east] at (-0.072,-1.000) {\texttt{two-way}};
    \draw[line width=\whiskLW, opacity=\whiskOp, draw=stroke1!70] (0.75000,-1.000) -- (1.52700,-1.000);
    \draw[line width=\whiskLW, opacity=\whiskOp, draw=stroke1!70] (0.75000,-1.085) -- (0.75000,-0.915);
    \draw[line width=\whiskLW, opacity=\whiskOp, draw=stroke1!70] (1.52700,-1.085) -- (1.52700,-0.915);
    \path[fill=stroke1!70, opacity=\boxOp] (1.03500,-1.160) rectangle (1.36100,-0.840);
    \draw[line width=\boxLW, draw=stroke1!70] (1.03500,-1.160) rectangle (1.36100,-0.840);
    \draw[line width=\medLW, draw=stroke1!70] (1.14050,-1.160) -- (1.14050,-0.840);
    \fill[stroke1!70] (1.17382,-1.000) circle (\dotR);
    \draw[line width=\baseLW, opacity=\whiskOp, draw=stroke1!70] (1.27800,-1.000) circle (\baseR);

    \node[anchor=east] at (-0.072,-2.000) {\texttt{three-way}};
    \draw[line width=\whiskLW, opacity=\whiskOp, draw=stroke2!70] (0.77800,-2.000) -- (1.61300,-2.000);
    \draw[line width=\whiskLW, opacity=\whiskOp, draw=stroke2!70] (0.77800,-2.085) -- (0.77800,-1.915);
    \draw[line width=\whiskLW, opacity=\whiskOp, draw=stroke2!70] (1.61300,-2.085) -- (1.61300,-1.915);
    \path[fill=stroke2!70, opacity=\boxOp] (1.09725,-2.160) rectangle (1.36875,-1.840);
    \draw[line width=\boxLW, draw=stroke2!70] (1.09725,-2.160) rectangle (1.36875,-1.840);
    \draw[line width=\medLW, draw=stroke2!70] (1.19400,-2.160) -- (1.19400,-1.840);
    \fill[stroke2!70] (1.22036,-2.000) circle (\dotR);
    \draw[line width=\baseLW, opacity=\whiskOp, draw=stroke2!70] (1.02800,-2.000) circle (\baseR);

\end{tikzpicture}

%% file: figures/fig_rq3_config_score_dist_boxplot_deepseek.tex
\begin{tikzpicture}[x=0.8888888888888887cm, y=0.75cm, font=\scriptsize]
    \definecolor{stroke2}{RGB}{230,159,0}
    \definecolor{stroke3}{RGB}{117,112,179}
    \draw[gray!55, solid] (0.000,0.550) -- (0.000,-2.550);
    \draw (0.000,-3.250) -- (0.000,-3.350);
    \node[anchor=north] at (0.000,-3.430) {0};
    \draw[gray!20, solid] (1.000,0.550) -- (1.000,-2.550);
    \draw (1.000,-3.250) -- (1.000,-3.350);
    \node[anchor=north] at (1.000,-3.430) {1};
    \draw[gray!20, solid] (2.000,0.550) -- (2.000,-2.550);
    \draw (2.000,-3.250) -- (2.000,-3.350);
    \node[anchor=north] at (2.000,-3.430) {2};
    \draw[gray!20, solid] (3.000,0.550) -- (3.000,-2.550);
    \draw (3.000,-3.250) -- (3.000,-3.350);
    \node[anchor=north] at (3.000,-3.430) {3};
    \draw[gray!20, solid] (4.000,0.550) -- (4.000,-2.550);
    \draw (4.000,-3.250) -- (4.000,-3.350);
    \node[anchor=north] at (4.000,-3.430) {4};
    \draw[gray!20, solid] (5.000,0.550) -- (5.000,-2.550);
    \draw (5.000,-3.250) -- (5.000,-3.350);
    \node[anchor=north] at (5.000,-3.430) {5};
    \draw[gray!20, solid] (6.000,0.550) -- (6.000,-2.550);
    \draw (6.000,-3.250) -- (6.000,-3.350);
    \node[anchor=north] at (6.000,-3.430) {6};
    \draw[->] (0.000,-3.250) -- (6.300,-3.250);
    \node[anchor=north] at (3.000,-3.800) {$E[\sum C_i]$};

    \def\boxHalf{0.160}\def\capHalf{0.085}\def\dotR{0.045}\def\baseR{0.060}
    \def\whiskLW{0.45pt}\def\whiskOp{0.90}\def\boxLW{0.45pt}\def\medLW{0.70pt}\def\baseLW{0.45pt}\def\boxOp{0.22}

    \node[anchor=east] at (-0.072,-0.000) {\texttt{isolated}};
    \draw[line width=\whiskLW, opacity=\whiskOp, draw=gray!35] (1.02900,-0.000) -- (2.55600,-0.000);
    \draw[line width=\whiskLW, opacity=\whiskOp, draw=gray!35] (1.02900,-0.085) -- (1.02900,0.085);
    \draw[line width=\whiskLW, opacity=\whiskOp, draw=gray!35] (2.55600,-0.085) -- (2.55600,0.085);
    \path[fill=gray!35, opacity=\boxOp] (1.43100,-0.160) rectangle (1.59750,0.160);
    \draw[line width=\boxLW, draw=gray!35] (1.43100,-0.160) rectangle (1.59750,0.160);
    \draw[line width=\medLW, draw=gray!35] (1.51350,-0.160) -- (1.51350,0.160);
    \fill[gray!35] (1.59050,-0.000) circle (\dotR);
    \draw[line width=\baseLW, opacity=\whiskOp, draw=gray!35] (4.52800,-0.000) circle (\baseR);

    \node[anchor=east] at (-0.072,-1.000) {\texttt{two-way}};
    \draw[line width=\whiskLW, opacity=\whiskOp, draw=stroke1!70] (1.08400,-1.000) -- (3.69500,-1.000);
    \draw[line width=\whiskLW, opacity=\whiskOp, draw=stroke1!70] (1.08400,-1.085) -- (1.08400,-0.915);
    \draw[line width=\whiskLW, opacity=\whiskOp, draw=stroke1!70] (3.69500,-1.085) -- (3.69500,-0.915);
    \path[fill=stroke1!70, opacity=\boxOp] (1.64600,-1.160) rectangle (2.70025,-0.840);
    \draw[line width=\boxLW, draw=stroke1!70] (1.64600,-1.160) rectangle (2.70025,-0.840);
    \draw[line width=\medLW, draw=stroke1!70] (2.05650,-1.160) -- (2.05650,-0.840);
    \fill[stroke1!70] (2.17989,-1.000) circle (\dotR);
    \draw[line width=\baseLW, opacity=\whiskOp, draw=stroke1!70] (4.80600,-1.000) circle (\baseR);

    \node[anchor=east] at (-0.072,-2.000) {\texttt{three-way}};
    \draw[line width=\whiskLW, opacity=\whiskOp, draw=stroke2!70] (1.30600,-2.000) -- (4.47200,-2.000);
    \draw[line width=\whiskLW, opacity=\whiskOp, draw=stroke2!70] (1.30600,-2.085) -- (1.30600,-1.915);
    \draw[line width=\whiskLW, opacity=\whiskOp, draw=stroke2!70] (4.47200,-2.085) -- (4.47200,-1.915);
    \path[fill=stroke2!70, opacity=\boxOp] (2.21525,-2.160) rectangle (3.39525,-1.840);
    \draw[line width=\boxLW, draw=stroke2!70] (2.21525,-2.160) rectangle (3.39525,-1.840);
    \draw[line width=\medLW, draw=stroke2!70] (2.77750,-2.160) -- (2.77750,-1.840);
    \fill[stroke2!70] (2.80957,-2.000) circle (\dotR);
    \draw[line width=\baseLW, opacity=\whiskOp, draw=stroke2!70] (4.52700,-2.000) circle (\baseR);

\end{tikzpicture}

%% file: figures/fig_rq3_config_score_dist_boxplot_grok.tex
\begin{tikzpicture}[x=0.8888888888888887cm, y=0.75cm, font=\scriptsize]
    \definecolor{stroke2}{RGB}{230,159,0}
    \definecolor{stroke3}{RGB}{117,112,179}
    \draw[gray!55, solid] (0.000,0.550) -- (0.000,-2.550);
    \draw (0.000,-3.250) -- (0.000,-3.350);
    \node[anchor=north] at (0.000,-3.430) {0};
    \draw[gray!20, solid] (1.000,0.550) -- (1.000,-2.550);
    \draw (1.000,-3.250) -- (1.000,-3.350);
    \node[anchor=north] at (1.000,-3.430) {1};
    \draw[gray!20, solid] (2.000,0.550) -- (2.000,-2.550);
    \draw (2.000,-3.250) -- (2.000,-3.350);
    \node[anchor=north] at (2.000,-3.430) {2};
    \draw[gray!20, solid] (3.000,0.550) -- (3.000,-2.550);
    \draw (3.000,-3.250) -- (3.000,-3.350);
    \node[anchor=north] at (3.000,-3.430) {3};
    \draw[gray!20, solid] (4.000,0.550) -- (4.000,-2.550);
    \draw (4.000,-3.250) -- (4.000,-3.350);
    \node[anchor=north] at (4.000,-3.430) {4};
    \draw[gray!20, solid] (5.000,0.550) -- (5.000,-2.550);
    \draw (5.000,-3.250) -- (5.000,-3.350);
    \node[anchor=north] at (5.000,-3.430) {5};
    \draw[gray!20, solid] (6.000,0.550) -- (6.000,-2.550);
    \draw (6.000,-3.250) -- (6.000,-3.350);
    \node[anchor=north] at (6.000,-3.430) {6};
    \draw[->] (0.000,-3.250) -- (6.300,-3.250);
    \node[anchor=north] at (3.000,-3.800) {$E[\sum C_i]$};

    \def\boxHalf{0.160}\def\capHalf{0.085}\def\dotR{0.045}\def\baseR{0.060}
    \def\whiskLW{0.45pt}\def\whiskOp{0.90}\def\boxLW{0.45pt}\def\medLW{0.70pt}\def\baseLW{0.45pt}\def\boxOp{0.22}

    \node[anchor=east] at (-0.072,-0.000) {\texttt{isolated}};
    \draw[line width=\whiskLW, opacity=\whiskOp, draw=gray!35] (1.08400,-0.000) -- (2.52800,-0.000);
    \draw[line width=\whiskLW, opacity=\whiskOp, draw=gray!35] (1.08400,-0.085) -- (1.08400,0.085);
    \draw[line width=\whiskLW, opacity=\whiskOp, draw=gray!35] (2.52800,-0.085) -- (2.52800,0.085);
    \path[fill=gray!35, opacity=\boxOp] (2.08425,-0.160) rectangle (2.42425,0.160);
    \draw[line width=\boxLW, draw=gray!35] (2.08425,-0.160) rectangle (2.42425,0.160);
    \draw[line width=\medLW, draw=gray!35] (2.25000,-0.160) -- (2.25000,0.160);
    \fill[gray!35] (2.13575,-0.000) circle (\dotR);
    \draw[line width=\baseLW, opacity=\whiskOp, draw=gray!35] (4.08200,-0.000) circle (\baseR);

    \node[anchor=east] at (-0.072,-1.000) {\texttt{two-way}};
    \draw[line width=\whiskLW, opacity=\whiskOp, draw=stroke1!70] (1.77800,-1.000) -- (3.38900,-1.000);
    \draw[line width=\whiskLW, opacity=\whiskOp, draw=stroke1!70] (1.77800,-1.085) -- (1.77800,-0.915);
    \draw[line width=\whiskLW, opacity=\whiskOp, draw=stroke1!70] (3.38900,-1.085) -- (3.38900,-0.915);
    \path[fill=stroke1!70, opacity=\boxOp] (2.47200,-1.160) rectangle (3.10450,-0.840);
    \draw[line width=\boxLW, draw=stroke1!70] (2.47200,-1.160) rectangle (3.10450,-0.840);
    \draw[line width=\medLW, draw=stroke1!70] (2.72300,-1.160) -- (2.72300,-0.840);
    \fill[stroke1!70] (2.73311,-1.000) circle (\dotR);
    \draw[line width=\baseLW, opacity=\whiskOp, draw=stroke1!70] (3.91600,-1.000) circle (\baseR);

    \node[anchor=east] at (-0.072,-2.000) {\texttt{three-way}};
    \draw[line width=\whiskLW, opacity=\whiskOp, draw=stroke2!70] (2.08400,-2.000) -- (4.13900,-2.000);
    \draw[line width=\whiskLW, opacity=\whiskOp, draw=stroke2!70] (2.08400,-2.085) -- (2.08400,-1.915);
    \draw[line width=\whiskLW, opacity=\whiskOp, draw=stroke2!70] (4.13900,-2.085) -- (4.13900,-1.915);
    \path[fill=stroke2!70, opacity=\boxOp] (2.75000,-2.160) rectangle (3.39575,-1.840);
    \draw[line width=\boxLW, draw=stroke2!70] (2.75000,-2.160) rectangle (3.39575,-1.840);
    \draw[line width=\medLW, draw=stroke2!70] (3.00050,-2.160) -- (3.00050,-1.840);
    \fill[stroke2!70] (3.06271,-2.000) circle (\dotR);
    \draw[line width=\baseLW, opacity=\whiskOp, draw=stroke2!70] (3.89000,-2.000) circle (\baseR);

\end{tikzpicture}

%% file: figures/fig_rq4_delta_minimality.tex
\resizebox{\linewidth}{!}{%
\begin{tikzpicture}[x=5.333333333333333cm, y=0.55cm, font=\footnotesize]
    \definecolor{stroke2}{RGB}{230,159,0}
    \def\cTwo{stroke1!70}
    \def\cThree{stroke2!70}

    \draw[gray!20, solid] (-0.600,0.550) -- (-0.600,-2.550);
    \draw (-0.600,-3.250) -- (-0.600,-3.350);
    \node[anchor=north] at (-0.600,-3.430) {-0.6};
    \draw[gray!20, solid] (-0.400,0.550) -- (-0.400,-2.550);
    \draw (-0.400,-3.250) -- (-0.400,-3.350);
    \node[anchor=north] at (-0.400,-3.430) {-0.4};
    \draw[gray!20, solid] (-0.200,0.550) -- (-0.200,-2.550);
    \draw (-0.200,-3.250) -- (-0.200,-3.350);
    \node[anchor=north] at (-0.200,-3.430) {-0.2};
    \draw[gray!55, solid] (-0.000,0.550) -- (-0.000,-2.550);
    \draw (-0.000,-3.250) -- (-0.000,-3.350);
    \node[anchor=north] at (-0.000,-3.430) {0};
    \draw[gray!20, solid] (0.200,0.550) -- (0.200,-2.550);
    \draw (0.200,-3.250) -- (0.200,-3.350);
    \node[anchor=north] at (0.200,-3.430) {0.2};
    \draw[gray!20, solid] (0.400,0.550) -- (0.400,-2.550);
    \draw (0.400,-3.250) -- (0.400,-3.350);
    \node[anchor=north] at (0.400,-3.430) {0.4};
    \draw[gray!20, solid] (0.600,0.550) -- (0.600,-2.550);
    \draw (0.600,-3.250) -- (0.600,-3.350);
    \node[anchor=north] at (0.600,-3.430) {0.6};
    \draw[->] (-0.600,-3.250) -- (0.660,-3.250);
    \node[anchor=north] at (0.000,-3.800) {$\Delta$};

    \def\dotR{0.045}\def\capHalf{0.085}\def\whiskLW{0.45pt}\def\whiskOp{0.90}
    \def\markSize{2.6pt}
    \def\drawwhisker#1#2#3#4#5{%
        \pgfmathsetmacro{\yLo}{#5-\capHalf}%
        \pgfmathsetmacro{\yHi}{#5+\capHalf}%
        \draw[#1, line width=\whiskLW, opacity=\whiskOp, draw=#2] (#3,#5) -- (#4,#5);%
        \draw[#1, line width=\whiskLW, opacity=\whiskOp, draw=#2] (#3,\yLo) -- (#3,\yHi);%
        \draw[#1, line width=\whiskLW, opacity=\whiskOp, draw=#2] (#4,\yLo) -- (#4,\yHi);%
    }


    \node[anchor=east] at (-0.614,-0.000) {\texttt{gpt-5-mini}};
    \drawwhisker{solid}{\cTwo}{-0.53181}{0.01230}{0.170}
    \node[circle, inner sep=0pt, outer sep=0pt, minimum size=\markSize, fill=\cTwo] at (-0.27239,0.170) {};
    \drawwhisker{solid}{\cThree}{-0.44112}{0.11839}{-0.170}
    \node[circle, inner sep=0pt, outer sep=0pt, minimum size=\markSize, draw=\cThree, fill=white, line width=\whiskLW] at (-0.18787,-0.170) {};

    \node[anchor=east] at (-0.614,-1.000) {\texttt{DeepSeek-V3.2}};
    \drawwhisker{solid}{\cTwo}{-0.37656}{0.30670}{-0.830}
    \node[circle, inner sep=0pt, outer sep=0pt, minimum size=\markSize, fill=\cTwo] at (-0.05649,-0.830) {};
    \drawwhisker{solid}{\cThree}{-0.13643}{0.51201}{-1.170}
    \node[circle, inner sep=0pt, outer sep=0pt, minimum size=\markSize, draw=\cThree, fill=white, line width=\whiskLW] at (0.19209,-1.170) {};

    \node[anchor=east] at (-0.614,-2.000) {\texttt{Grok-4.1-fast}};
    \drawwhisker{solid}{\cTwo}{-0.32159}{0.35898}{-1.830}
    \node[circle, inner sep=0pt, outer sep=0pt, minimum size=\markSize, fill=\cTwo] at (-0.01189,-1.830) {};
    \drawwhisker{solid}{\cThree}{-0.36794}{0.47979}{-2.170}
    \node[circle, inner sep=0pt, outer sep=0pt, minimum size=\markSize, draw=\cThree, fill=white, line width=\whiskLW] at (0.04059,-2.170) {};
\end{tikzpicture}%
}

%% file: figures/fig_rq4_delta_line_deviation.tex
\resizebox{\linewidth}{!}{%
\begin{tikzpicture}[x=0.05333333333333333cm, y=0.55cm, font=\scriptsize]
    \definecolor{stroke2}{RGB}{230,159,0}
    \def\cTwo{stroke1!70}
    \def\cThree{stroke2!70}

    \draw[gray!20, solid] (-60.000,0.550) -- (-60.000,-2.550);
    \draw (-60.000,-3.250) -- (-60.000,-3.350);
    \node[anchor=north] at (-60.000,-3.430) {-6};
    \draw[gray!20, solid] (-40.000,0.550) -- (-40.000,-2.550);
    \draw (-40.000,-3.250) -- (-40.000,-3.350);
    \node[anchor=north] at (-40.000,-3.430) {-4};
    \draw[gray!20, solid] (-20.000,0.550) -- (-20.000,-2.550);
    \draw (-20.000,-3.250) -- (-20.000,-3.350);
    \node[anchor=north] at (-20.000,-3.430) {-2};
    \draw[gray!55, solid] (0.000,0.550) -- (0.000,-2.550);
    \draw (0.000,-3.250) -- (0.000,-3.350);
    \node[anchor=north] at (0.000,-3.430) {0};
    \draw[gray!20, solid] (20.000,0.550) -- (20.000,-2.550);
    \draw (20.000,-3.250) -- (20.000,-3.350);
    \node[anchor=north] at (20.000,-3.430) {2};
    \draw[gray!20, solid] (40.000,0.550) -- (40.000,-2.550);
    \draw (40.000,-3.250) -- (40.000,-3.350);
    \node[anchor=north] at (40.000,-3.430) {4};
    \draw[->] (-60.000,-3.250) -- (45.000,-3.250);
    \node[anchor=north] at (0.000,-3.800) {$\Delta$};

    \def\dotR{0.045}\def\capHalf{0.085}\def\whiskLW{0.45pt}\def\whiskOp{0.90}
    \def\markSize{2.6pt}
    \def\drawwhisker#1#2#3#4#5{%
        \pgfmathsetmacro{\yLo}{#5-\capHalf}%
        \pgfmathsetmacro{\yHi}{#5+\capHalf}%
        \draw[#1, line width=\whiskLW, opacity=\whiskOp, draw=#2] (#3,#5) -- (#4,#5);%
        \draw[#1, line width=\whiskLW, opacity=\whiskOp, draw=#2] (#3,\yLo) -- (#3,\yHi);%
        \draw[#1, line width=\whiskLW, opacity=\whiskOp, draw=#2] (#4,\yLo) -- (#4,\yHi);%
    }

    \node[anchor=east] at (-61.200,-0.000) {\texttt{gpt-5-mini}};
    \drawwhisker{solid}{\cTwo}{-14.23357}{20.02285}{0.170}
    \node[circle, inner sep=0pt, outer sep=0pt, minimum size=\markSize, fill=\cTwo] at (5.09417,0.170) {};
    \drawwhisker{solid}{\cThree}{-29.29896}{4.57863}{-0.170}
    \node[circle, inner sep=0pt, outer sep=0pt, minimum size=\markSize, draw=\cThree, fill=white, line width=\whiskLW] at (-5.14051,-0.170) {};

    \node[anchor=east] at (-61.200,-1.000) {\texttt{DeepSeek-V3.2}};
    \drawwhisker{solid}{\cTwo}{-39.54000}{3.04325}{-0.830}
    \node[circle, inner sep=0pt, outer sep=0pt, minimum size=\markSize, fill=\cTwo] at (-9.03486,-0.830) {};
    \drawwhisker{solid}{\cThree}{-53.82025}{-0.86587}{-1.170}
    \node[circle, inner sep=0pt, outer sep=0pt, minimum size=\markSize, draw=\cThree, fill=white, line width=\whiskLW] at (-24.64401,-1.170) {};

    \node[anchor=east] at (-61.200,-2.000) {\texttt{Grok-4.1-fast}};
    \drawwhisker{solid}{\cTwo}{-21.83545}{9.87338}{-1.830}
    \node[circle, inner sep=0pt, outer sep=0pt, minimum size=\markSize, fill=\cTwo] at (-2.27693,-1.830) {};
    \drawwhisker{solid}{\cThree}{-44.77842}{1.85091}{-2.170}
    \node[circle, inner sep=0pt, outer sep=0pt, minimum size=\markSize, draw=\cThree, fill=white, line width=\whiskLW] at (-14.69745,-2.170) {};
\end{tikzpicture}%
}

%% file: figures/fig_rq4_delta_levenshtein.tex
\resizebox{\linewidth}{!}{%
\begin{tikzpicture}[x=5.333333333333333cm, y=0.55cm, font=\scriptsize]
    \definecolor{stroke2}{RGB}{230,159,0}
    \def\cTwo{stroke1!70}
    \def\cThree{stroke2!70}

    \draw[gray!20, solid] (-0.600,0.550) -- (-0.600,-2.550);
    \draw (-0.600,-3.250) -- (-0.600,-3.350);
    \node[anchor=north] at (-0.600,-3.430) {-0.6};
    \draw[gray!20, solid] (-0.400,0.550) -- (-0.400,-2.550);
    \draw (-0.400,-3.250) -- (-0.400,-3.350);
    \node[anchor=north] at (-0.400,-3.430) {-0.4};
    \draw[gray!20, solid] (-0.200,0.550) -- (-0.200,-2.550);
    \draw (-0.200,-3.250) -- (-0.200,-3.350);
    \node[anchor=north] at (-0.200,-3.430) {-0.2};
    \draw[gray!55, solid] (0.000,0.550) -- (0.000,-2.550);
    \draw (0.000,-3.250) -- (0.000,-3.350);
    \node[anchor=north] at (0.000,-3.430) {0};
    \draw[gray!20, solid] (0.200,0.550) -- (0.200,-2.550);
    \draw (0.200,-3.250) -- (0.200,-3.350);
    \node[anchor=north] at (0.200,-3.430) {0.2};
    \draw[gray!20, solid] (0.400,0.550) -- (0.400,-2.550);
    \draw (0.400,-3.250) -- (0.400,-3.350);
    \node[anchor=north] at (0.400,-3.430) {0.4};
    \draw[gray!20, solid] (0.600,0.550) -- (0.600,-2.550);
    \draw (0.600,-3.250) -- (0.600,-3.350);
    \node[anchor=north] at (0.600,-3.430) {0.6};
    \draw[->] (-0.600,-3.250) -- (0.660,-3.250);
    \node[anchor=north] at (0.000,-3.800) {$\Delta$};

    \def\dotR{0.045}\def\capHalf{0.085}\def\whiskLW{0.45pt}\def\whiskOp{0.90}
    \def\markSize{2.6pt}
    \def\drawwhisker#1#2#3#4#5{%
        \pgfmathsetmacro{\yLo}{#5-\capHalf}%
        \pgfmathsetmacro{\yHi}{#5+\capHalf}%
        \draw[#1, line width=\whiskLW, opacity=\whiskOp, draw=#2] (#3,#5) -- (#4,#5);%
        \draw[#1, line width=\whiskLW, opacity=\whiskOp, draw=#2] (#3,\yLo) -- (#3,\yHi);%
        \draw[#1, line width=\whiskLW, opacity=\whiskOp, draw=#2] (#4,\yLo) -- (#4,\yHi);%
    }

    \node[anchor=east] at (-0.614,-0.000) {\texttt{gpt-5-mini}};
    \drawwhisker{solid}{\cTwo}{-0.12776}{0.22216}{0.170}
    \node[circle, inner sep=0pt, outer sep=0pt, minimum size=\markSize, fill=\cTwo] at (0.10021,0.170) {};
    \drawwhisker{solid}{\cThree}{-0.25379}{0.12264}{-0.170}
    \node[circle, inner sep=0pt, outer sep=0pt, minimum size=\markSize, draw=\cThree, fill=white, line width=\whiskLW] at (-0.01400,-0.170) {};

    \node[anchor=east] at (-0.614,-1.000) {\texttt{DeepSeek-V3.2}};
    \drawwhisker{solid}{\cTwo}{-0.33539}{0.01632}{-0.830}
    \node[circle, inner sep=0pt, outer sep=0pt, minimum size=\markSize, fill=\cTwo] at (-0.09804,-0.830) {};
    \drawwhisker{solid}{\cThree}{-0.44382}{-0.02624}{-1.170}
    \node[circle, inner sep=0pt, outer sep=0pt, minimum size=\markSize, draw=\cThree, fill=white, line width=\whiskLW] at (-0.22943,-1.170) {};

    \node[anchor=east] at (-0.614,-2.000) {\texttt{Grok-4.1-fast}};
    \drawwhisker{solid}{\cTwo}{-0.24113}{0.07780}{-1.830}
    \node[circle, inner sep=0pt, outer sep=0pt, minimum size=\markSize, fill=\cTwo] at (-0.04311,-1.830) {};
    \drawwhisker{solid}{\cThree}{-0.39858}{0.01714}{-2.170}
    \node[circle, inner sep=0pt, outer sep=0pt, minimum size=\markSize, draw=\cThree, fill=white, line width=\whiskLW] at (-0.14159,-2.170) {};
\end{tikzpicture}%
}

%% file: figures/fig_rq4_delta_spurious.tex
\resizebox{\linewidth}{!}{%
\begin{tikzpicture}[x=0.6666666666666666cm, y=0.55cm, font=\scriptsize]
    \definecolor{stroke2}{RGB}{230,159,0}
    \def\cTwo{stroke1!70}
    \def\cThree{stroke2!70}

    \draw[gray!20, solid] (-4.000,0.550) -- (-4.000,-2.550);
    \draw (-4.000,-3.250) -- (-4.000,-3.350);
    \node[anchor=north] at (-4.000,-3.430) {-4};
    \draw[gray!20, solid] (-2.000,0.550) -- (-2.000,-2.550);
    \draw (-2.000,-3.250) -- (-2.000,-3.350);
    \node[anchor=north] at (-2.000,-3.430) {-2};
    \draw[gray!55, solid] (0.000,0.550) -- (0.000,-2.550);
    \draw (0.000,-3.250) -- (0.000,-3.350);
    \node[anchor=north] at (0.000,-3.430) {0};
    \draw[gray!20, solid] (2.000,0.550) -- (2.000,-2.550);
    \draw (2.000,-3.250) -- (2.000,-3.350);
    \node[anchor=north] at (2.000,-3.430) {2};
    \draw[gray!20, solid] (4.000,0.550) -- (4.000,-2.550);
    \draw (4.000,-3.250) -- (4.000,-3.350);
    \node[anchor=north] at (4.000,-3.430) {4};
    \draw[->] (-4.000,-3.250) -- (4.500,-3.250);
    \node[anchor=north] at (0.000,-3.800) {$\Delta$};

    \def\dotR{0.045}\def\capHalf{0.085}\def\whiskLW{0.45pt}\def\whiskOp{0.90}
    \def\markSize{2.6pt}
    \def\drawwhisker#1#2#3#4#5{%
        \pgfmathsetmacro{\yLo}{#5-\capHalf}%
        \pgfmathsetmacro{\yHi}{#5+\capHalf}%
        \draw[#1, line width=\whiskLW, opacity=\whiskOp, draw=#2] (#3,#5) -- (#4,#5);%
        \draw[#1, line width=\whiskLW, opacity=\whiskOp, draw=#2] (#3,\yLo) -- (#3,\yHi);%
        \draw[#1, line width=\whiskLW, opacity=\whiskOp, draw=#2] (#4,\yLo) -- (#4,\yHi);%
    }

    \node[anchor=east] at (-4.080,-0.000) {\texttt{gpt-5-mini}};
    \drawwhisker{solid}{\cTwo}{-1.60810}{3.78238}{0.170}
    \node[circle, inner sep=0pt, outer sep=0pt, minimum size=\markSize, fill=\cTwo] at (1.06631,0.170) {};
    \drawwhisker{solid}{\cThree}{-1.69931}{2.24408}{-0.170}
    \node[circle, inner sep=0pt, outer sep=0pt, minimum size=\markSize, draw=\cThree, fill=white, line width=\whiskLW] at (0.48013,-0.170) {};

    \node[anchor=east] at (-4.080,-1.000) {\texttt{DeepSeek-V3.2}};
    \drawwhisker{solid}{\cTwo}{-2.43658}{0.43621}{-0.830}
    \node[circle, inner sep=0pt, outer sep=0pt, minimum size=\markSize, fill=\cTwo] at (-0.52956,-0.830) {};
    \drawwhisker{solid}{\cThree}{-2.81052}{0.67623}{-1.170}
    \node[circle, inner sep=0pt, outer sep=0pt, minimum size=\markSize, draw=\cThree, fill=white, line width=\whiskLW] at (-0.91019,-1.170) {};

    \node[anchor=east] at (-4.080,-2.000) {\texttt{Grok-4.1-fast}};
    \drawwhisker{solid}{\cTwo}{-1.44464}{0.60683}{-1.830}
    \node[circle, inner sep=0pt, outer sep=0pt, minimum size=\markSize, fill=\cTwo] at (-0.13344,-1.830) {};
    \drawwhisker{solid}{\cThree}{-1.48241}{0.82213}{-2.170}
    \node[circle, inner sep=0pt, outer sep=0pt, minimum size=\markSize, draw=\cThree, fill=white, line width=\whiskLW] at (-0.18917,-2.170) {};
\end{tikzpicture}%
}

%% file: figures/fig_rq4_delta_delta_volume.tex
\resizebox{\linewidth}{!}{%
\begin{tikzpicture}[x=0.05333333333333333cm, y=0.55cm, font=\scriptsize]
    \definecolor{stroke2}{RGB}{230,159,0}
    \def\cTwo{stroke1!70}
    \def\cThree{stroke2!70}

    \draw[gray!20, solid] (-40.000,0.550) -- (-40.000,-2.550);
    \draw (-40.000,-3.250) -- (-40.000,-3.350);
    \node[anchor=north] at (-40.000,-3.430) {-4};
    \draw[gray!20, solid] (-20.000,0.550) -- (-20.000,-2.550);
    \draw (-20.000,-3.250) -- (-20.000,-3.350);
    \node[anchor=north] at (-20.000,-3.430) {-2};
    \draw[gray!55, solid] (0.000,0.550) -- (0.000,-2.550);
    \draw (0.000,-3.250) -- (0.000,-3.350);
    \node[anchor=north] at (0.000,-3.430) {0};
    \draw[gray!20, solid] (20.000,0.550) -- (20.000,-2.550);
    \draw (20.000,-3.250) -- (20.000,-3.350);
    \node[anchor=north] at (20.000,-3.430) {2};
    \draw[gray!20, solid] (40.000,0.550) -- (40.000,-2.550);
    \draw (40.000,-3.250) -- (40.000,-3.350);
    \node[anchor=north] at (40.000,-3.430) {4};
    \draw[gray!20, solid] (60.000,0.550) -- (60.000,-2.550);
    \draw (60.000,-3.250) -- (60.000,-3.350);
    \node[anchor=north] at (60.000,-3.430) {6};
    \draw[->] (-40.000,-3.250) -- (65.000,-3.250);
    \node[anchor=north] at (0.000,-3.800) {$\Delta$};

    \def\dotR{0.045}\def\capHalf{0.085}\def\whiskLW{0.45pt}\def\whiskOp{0.90}
    \def\markSize{2.6pt}
    \def\drawwhisker#1#2#3#4#5{%
        \pgfmathsetmacro{\yLo}{#5-\capHalf}%
        \pgfmathsetmacro{\yHi}{#5+\capHalf}%
        \draw[#1, line width=\whiskLW, opacity=\whiskOp, draw=#2] (#3,#5) -- (#4,#5);%
        \draw[#1, line width=\whiskLW, opacity=\whiskOp, draw=#2] (#3,\yLo) -- (#3,\yHi);%
        \draw[#1, line width=\whiskLW, opacity=\whiskOp, draw=#2] (#4,\yLo) -- (#4,\yHi);%
    }

    \node[anchor=east] at (-41.200,-0.000) {\texttt{gpt-5-mini}};
    \drawwhisker{solid}{\cTwo}{-24.62885}{20.26455}{0.170}
    \node[circle, inner sep=0pt, outer sep=0pt, minimum size=\markSize, fill=\cTwo] at (1.42201,0.170) {};
    \drawwhisker{solid}{\cThree}{-24.70303}{25.06971}{-0.170}
    \node[circle, inner sep=0pt, outer sep=0pt, minimum size=\markSize, draw=\cThree, fill=white, line width=\whiskLW] at (3.12974,-0.170) {};

    \node[anchor=east] at (-41.200,-1.000) {\texttt{DeepSeek-V3.2}};
    \drawwhisker{solid}{\cTwo}{-1.87331}{45.94789}{-0.830}
    \node[circle, inner sep=0pt, outer sep=0pt, minimum size=\markSize, fill=\cTwo] at (19.00227,-0.830) {};
    \drawwhisker{solid}{\cThree}{-8.25839}{23.47367}{-1.170}
    \node[circle, inner sep=0pt, outer sep=0pt, minimum size=\markSize, draw=\cThree, fill=white, line width=\whiskLW] at (6.00913,-1.170) {};

    \node[anchor=east] at (-41.200,-2.000) {\texttt{Grok-4.1-fast}};
    \drawwhisker{solid}{\cTwo}{3.61463}{41.62761}{-1.830}
    \node[circle, inner sep=0pt, outer sep=0pt, minimum size=\markSize, fill=\cTwo] at (17.72253,-1.830) {};
    \drawwhisker{solid}{\cThree}{-14.23791}{28.52593}{-2.170}
    \node[circle, inner sep=0pt, outer sep=0pt, minimum size=\markSize, draw=\cThree, fill=white, line width=\whiskLW] at (3.94263,-2.170) {};
\end{tikzpicture}%
}

%% file: 6_discussion_and_implications/discussion_and_implications.tex
\section{Discussion: Implications and Limitations}

Failure-explanation quality emerges from the interaction between model behavior, context composition, and evaluation criteria. Our results show that models often exploit the same evidence in different ways. Causal and action-oriented criteria matter more for downstream repair than surface-level factors. LLM-as-a-judge can approximate human judgments on these dimensions, but this raises questions about how far context engineering and judge calibration can push explanation quality without additional model training.

\subsection{Implications}
Failure explanations must be grounded in actionable evidence. In our study, \texttt{CODE} and \texttt{TEST} most strongly support causal and action-\allowbreak{}oriented criteria for \texttt{DeepSeek\allowbreak-V3.2} and \texttt{Grok\allowbreak-4.1-fast}, while \texttt{gpt\allowbreak-5-mini} benefits more from slice-derived modules (e.g., \texttt{SLICE\_UNION}) that primarily strengthen grounding (C5). In composed settings, \texttt{DOCSTRING} often has negative marginal effects, suggesting it can add noise relative to evidence that constrains the failure mechanism. Context partitioning makes these tradeoffs explicit, enabling systems to compose multi-source evidence under a context budget instead of assuming \textit{more-context-is-better}~\cite{Liu2023LostInMiddle}.

These observations imply that context selection should be explicit: systems should prioritize executable artifacts, then add slices when they contribute relevant signals. Because the best strategy is model-dependent, context composition should remain configurable and potentially learnable.

Similarly, evaluators should not collapse explanation quality into a single proxy metric. Reporting should separate causal understanding, actionability, grounding, and presentation, and treat LLM-as-a-judge as a measurement instrument that must be validated against humans.

%% file: 7_threats_to_validity/threats_to_validity.tex
\subsection{Limitations: Threats to Validity}

\textbf{Internal Validity} -- Observed effects may partly reflect execution variance, service-level effects, and backend differences in structured-output enforcement rather than configuration content; LLM outputs also remain non-deterministic despite our determinism controls and three repeated runs per model. Our repair protocol is applied uniformly, but function-only splicing and triggering-test validation can still introduce measurement noise, so rankings among similar configurations should be interpreted with these threats in mind.

\textbf{External Validity} -- Our defect set is intentionally narrow: 12 localized, reproducible Python defects translated from Defects4J. Translation can introduce artifacts that affect context extraction and model behavior, so observed effects should be interpreted as directional and not assumed to transfer unchanged to native Python projects or back to the original Java defects. Larger benchmarks are needed to assess stability across projects, failure types, model families, and context taxonomies.

\textbf{Construct Validity} -- Our six binary criteria (C1--C6) separate causal understanding, actionability, grounding, and presentation, but binarization compresses nuance, C1/C5 are heuristic proxies, and the unweighted sum treats all criteria as equally important. LLM-as-a-judge criteria (C2/C3/C4/C6) inherit judge-model biases and show weaker agreement with humans for brevity (C6). Downstream repair is also an imperfect utility proxy because validation reruns only the triggering test and reference-minimal-fix comparison depends on a single reference patch plus AST normalization.

\textbf{Statistical Conclusion Validity} -- Statistical conclusion validity is limited by the structure and size of the configuration batches, especially the isolated batch with only eight non-baseline configurations per model. We therefore treat isolated results descriptively; larger configuration sets and more repetitions would be needed for finer-grained inference.

%% file: 8_conclusion_and_future_work/conclusion_and_future_work.tex
\section{Conclusion and Future Work}
\textbf{Conclusion} -- 
We systematically evaluate LLM-\allowbreak{}generated failure explanations via \emph{context partitioning}, treating available artifacts as composable context modules. We operationalize explanation quality through six criteria (C1--C6) and find that LLM-as-a-judge aligns closely with human judgment on causal and action-oriented criteria (C2--C4), but less so on brevity (C6). Across three LLMs, executable evidence (especially \texttt{CODE} and \texttt{TEST}) strengthens quality for \texttt{DeepSeek\allowbreak-V3.2} and \texttt{Grok\allowbreak-4.1-fast}, while \texttt{gpt\allowbreak-5-mini} benefits more from slice-derived modules and \texttt{DOCSTRING} often adds noise. Higher explanation-score quartiles show higher downstream repair pass rates and, for some models, more reference-like fixes; low explanation-score quartiles can underperform the no-explanation baseline. These results support model-aware context-selection strategies.

\textbf{Future Work} -- We plan to automate per-failure context selection, handle multi-file defects, calibrate low-agreement criteria and judge bias, study developer-centric utility, examine ordering and formatting effects, and strengthen robustness across additional model families and deployment settings